\documentclass[preprint]{aastex}
\usepackage{graphicx}
\begin{document}

\title{Planet migration and gap formation by tidally-induced shocks.}
\author{R. R. Rafikov}
\affil{Princeton University Observatory, Princeton, NJ 08544}
\email{rrr@astro.princeton.edu}

\begin{abstract}
Gap formation in a gas disk triggered by 
disk-planet tidal interaction is considered.
Density waves launched by the planet  are assumed to be damped
 as a result of their nonlinear
evolution leading to shock formation and its subsequent dissipation.
As a consequence wave angular momentum is transferred to the disk, 
leading to evolution of its surface density.
Planetary migration is an important ingredient of the theory; 
effects of the planet-induced surface density perturbations
on the migration speed 
are considered. A gap is assumed 
to form when a stationary solution for the surface density profile is
no longer possible in the frame of reference migrating with the planet. 
An analytical
limit on the planetary mass necessary to open a gap in an inviscid 
disk is derived. 
The critical mass turns out to be smaller than mass $M_1$ for which
planetary Hill's radius equals disk scaleheight by a factor of at least
$Q^{5/7}$ ($Q$ is the Toomre stability parameter) depending on the strength
of the migration feedback.
In viscous disks the  critical planetary mass could vary
from $\sim 0.1M_1$ to $M_1$, depending on the disk viscosity. 
This implies that
a gap could be formed by a planet with mass $(1-10)~M_\oplus$ depending
on the disk aspect ratio, viscosity, and planet's location in the nebula.
\end{abstract}


\section{Introduction.}  \label{intro}

The recent discovery of extrasolar planets on orbits very close to their 
parent stars (Mayor \& Queloz 1995; Marcy et al. 2000; Vogt et al. 2000; 
Butler 2001) has raised a number of questions about the formation
mechanisms of such systems. These close-in planetary companions are
all presumed to be gas giants, typically with masses of the order of Jupiter 
mass $M_J\approx 2\times 10^{30}$ g. It is very unlikely that such 
planets were formed at their present locations (Boss 1995): current theories 
(Mizuno 1980; Bodenheimer \& Pollack 1986) predict that giant planets 
were formed by gas accretion onto massive
($\sim 15~M_\oplus$) rocky core
 which themselves are the result of
accumulation of a large number of icy planetesimals.  
The most favorable conditions for this process exist beyond 
the so-called ``snow line'' (Hayashi 1981; Sasselov \& Lecar 2000)
which is estimated  to lie several AU from the star, far larger than 
the actual orbital radii of the known extrasolar planets.

The most popular theory explaining this paradox presumes that giant planets
were indeed formed far outside 
their present locations but then migrated inwards
as a result of tidal interaction with a gaseous nebula (Ward 1997a).
If a planet is fully immersed in a gas disk then 
the migration process works as follows:
planetary gravitational torques produce density waves carrying 
angular momentum in the surrounding gas disk. 
For the planet to migrate inwards (outwards) 
two conditions must be fulfilled: (1) interaction with the 
outer (inner) part of
 the disk should be stronger than with the inner (outer) one, and (2),
the waves must not return to the planet. In a finite disk the only 
way to fulfill the second condition is for the waves to
dissipate so that their angular momentum gets transferred to the disk flow.
It was demonstrated by Ward (1986) that in Keplerian disks conditions are 
usually such that the planet tends to migrate inwards. The 
typical timescale
of this process, $\sim 10^3$ yr for a Jupiter-mass planet, is very short
compared with the lifetime of the nebula itself ($10^6-10^7$ yr). This 
short timescale
presents a significant problem for the migration scenario
because the migrating planet is likely to drift right into its parent
star in very short time. 

Various mechanisms to {\it stop} migration
have been proposed: formation of a cavity in the 
inner disk by magnetospheric activity of the central star
(Shu et al. 1994), 
resonant interaction with another
planet in the system (Masset \& Snellgrove 2001), 
spin-orbital interaction with the central star (Trilling et al. 1998), etc. 
On the other hand, 
there is  a process almost inevitable in the course of
giant planet formation --- gap formation in the disk near the planet
--- which can effectively {\it slow down} planet migration. 
When the planet is not very massive its gravity cannot strongly affect 
the disk surface density distribution in its vicinity and strong 
interaction with the gas disk through  Lindblad resonances 
(Goldreich \& Tremaine 1980, hereafter GT80) leads to the rapid migration
described above 
(so called ``type I migration'', Ward 1997a).  
As the planetary mass grows the drift speed increases but at some point
the strength of the torque exerted on the surrounding gas becomes so large 
that planet simply repels
the gas and a gap forms. 
This strongly diminishes the tidal interaction because
it is usually dominated by high-order Lindblad resonances which now lie inside 
the gap. Thus the orbital evolution of the planet becomes 
tied to the evolution of 
the disk and it migrates  on the viscous timescale of the disk which could
be much  longer ($10^5-10^6$ yr or even longer, 
depending on the viscosity in the disk). This stage is called ``type II 
migration'' and its existence might
help to explain the survival of planetary 
systems in the course of their orbital evolution or at least significantly 
alleviate this problem (Ward 1997). 
Gap formation also provides a reason for the existence of a maximum mass
of extrasolar giant planets (Nelson et al. 2000).

In view of all this a very important question arises: how massive should a
planet be in order to open a gap? The answer depends not only on the 
conditions in the nebula (surface density, temperature, viscosity)
but also on the dissipation mechanism of planet-induced density waves
and the mobility of the planet.
If the viscosity in the disk is absent and the planet is not drifting
there is no mechanism which could oppose tidal torques and a gap is opened
by an arbitrarily small perturber. This conclusion changes if planetary 
migration is taken into account self-consistently because a low-mass planet
is able to drift through the gap before gap fully forms (Ward \& Hourigan 
1989, hereafter WH89). The planet must slow down its migration 
somehow for the gap to be opened.

In their study WH89 demonstrated that the
disk surface density is enhanced in front of 
the drifting planet and is reduced behind it. 
This effect {\it diminishes} the difference between
the torques produced by the planet in the disk outside and inside its
orbit and acts to {\it slow down} 
the migration. For some mass of the planet its steady drift becomes 
impossible, the 
planet halts, and gap formation ensues. Thus, in this picture there is a 
minimum mass for gap opening even in an inviscid disk. Assuming that wave 
dissipation is a very rapid process and that
angular momentum is immediately
transferred to the disk WH89 estimated this critical 
mass to be
\begin{equation}
M_2\sim \frac{h}{r}M_f, ~~~\mbox{where}~~~M_f=\Sigma h^2.\label{MWH}
\end{equation}
Here $\Sigma$ is a surface density of the disk, $h\equiv c/\Omega$ is its 
geometric thickness (here $c$ is the sound speed and $\Omega$ 
is the angular orbital frequency), 
and $r$ is the distance from the central star.
Typical protosolar nebular parameters were summarized by Hayashi (1980) in the 
form of the minimum mass Solar nebula (MMSN) model:
\begin{eqnarray}
\Sigma_0(r)=1700~ \mbox{g cm}^{-3}\left(\frac{r}{1~\mbox{AU}}
\right)^{-3/2},~~~~ 
c_0(r)=1.2~ \mbox{km s}^{-1}\left(\frac{r}
{1~\mbox{AU}}\right)^{-1/4}.\label{hayashi}
\end{eqnarray}
 Using (\ref{hayashi}) we obtain that $M_f\approx 6\times 10^{26}\mbox{g}
\approx 0.1~M_\oplus$ and $M_2\sim 3\times 10^{25}$ g at $1$ AU 
which is about
the mass of Mercury. Of course, if the disk possesses nonzero viscosity 
this mass
would increase because viscous diffusion opposes gap formation.

Lin \& Papaloizou (1993, hereafter LP93) present a different point of view
on the planetary mass required to open a gap. 
They also assume damping of the density
waves to be local but require mass of the planet to be high enough for the 
waves to be strongly
 nonlinear from the beginning and shock immediately
transferring angular momentum to the fluid. They showed that this minimum mass
corresponds to the case when the Hill's  radius of the planet 
$R_H=r(M_p/M_\star)^{1/3}$ ($M_p$ is the mass of the planet, 
$M_\star$ is the mass of the central star) is comparable to the vertical disk
scaleheight yielding 
\begin{equation}
M_1=\frac{2c^3}{3\Omega G}\approx 14~M_\oplus\left(\frac{r}{1~\mbox{AU}}
\right)^{3/4}\label{M1}
\end{equation}
as the relevant mass for opening a gap 
[the numerical estimate in equation (\ref{M1}) 
is made for MMSN parameters given in equation 
(\ref{hayashi})]. 
Another way to look at this restriction is to notice that at $M_p\sim M_1$
the pressure gradient in the disk in the vicinity of the planet becomes so high
that epicyclic frequency becomes imaginary and Rayleigh's instability
develops. Thus, this restriction could be considered as an upper limit 
on the mass of the planet that does not open a gap.

Values of the gap opening mass estimated by these two approaches 
differ by a huge factor. Indeed:
\begin{equation}
\frac{M_1}{M_f}=\frac{2\pi}{3}\frac{\kappa c}{\pi G\Sigma}
=\frac{2\pi}{3}Q,\label{mass_rat}
\end{equation}
where $Q$ is the Toomre stability parameter (Binney \& Tremaine 1987) 
and $\kappa$ is the epicyclic frequency
($\kappa=\Omega$ for Keplerian rotation law); using MMSN parameters given 
by equation (\ref{hayashi}) we obtain that $Q\approx 70$  and $h/r\approx
0.04$ at $1$ AU
meaning that $M_2\sim (h/r)Q^{-1} M_1\approx 5\times 10^{-4} M_1$.

Ward \& Hourigan (WH89) assume that the damping mechanism is independent of 
the planetary mass and essentially linear, such as viscous dissipation 
(Takeuchi et al. 1996) or radiative damping (Cassen \& Woolum 1996). 
However these mechanisms are probably ineffective in cold, weakly ionized 
and optically thick systems such as protoplanetary disks 
(Hawley \& Stone 1998; Goodman \& Rafikov 2001, hereafter GR01); 
in this case tides raised 
by low-mass planets could propagate much further than just a fraction
of disk scale length and gap opening requires significantly 
more massive perturbers (Ward 1997a). 

Requiring waves to
be strongly nonlinear and damp immediately 
as a necessary condition for gap formation is also probably too radical. 
Indeed, density waves produced by the planet can still evolve due to
weak nonlinearity and are able to 
form a weak shock even if the planet is less massive
than $M_1$ (GR01). This mechanism can lead to a gap formation by lower mass
planets than LP93 assumed. Clearly to obtain a reliable estimate 
of the critical planetary mass one must use a realistic wave damping
prescription. 

Goodman \& Rafikov (GR01) have considered nonlinear evolution of
the density waves produced by low-mass planets in two-dimensional disks
using the shearing sheet approximation and assuming the background surface 
density and sound speed to be constant.  
They have found that a shock is formed quite rapidly (depending on $M_p/M_1$, 
typically
after travelling several disk scaleheights from the planet
but still not immediately)
because the radial wavelength of the perturbation constantly decreases while 
its amplitude increases (as a consequence of angular momentum flux 
conservation) thus facilitating wavefront breaking. After wave shocks
its angular momentum is gradually transferred to the disk fluid leading
to surface density evolution (thus violating the constant surface density 
assumption). Rafikov (2001) generalized this analysis 
 by taking  into consideration
the effects of spatial variations of the surface 
density and sound speed in the disk as well as the cylindrical geometry of
the problem. Realistic prescription of the 
global angular momentum dissipation was provided 
under the condition that the surface density and sound speed vary on scales
larger than the wavelength of the perturbation.

In this paper we study the criterion for  gap opening by planetary tides
by determining
under which conditions a steady-state solution for the 
disk surface density perturbation is no longer possible
(method used in WH89). Planet migration 
is taken into account self-consistently.
Tidal perturbations are supposed to be damped nonlocally 
 by weak nonlinearity and their angular momentum is assumed to be 
transferred to the disk, as described quantitatively in Rafikov (2001).
For the gap opening the assumption of varying surface density 
in Rafikov (2001) is the most 
important one because it allows us to solve the problem self-consistently
(angular momentum transfer depends on how the surface density 
is distributed radially). 

After developing the general analytical apparatus in \S \ref{geneq}
we first explore the case of an inviscid disk in \S \ref{visc-less}
to highlight the most important physical mechanisms relevant for gap
formation. Then we generalize consideration to include the disk viscosity
in \S \ref{viscous}. Finally, in \S \ref{disc} we discuss our results.

\section{Basic equations}\label{geneq}

We study the surface density evolution of a gas disk driven by 
planetary gravitational perturbations. In our model a planet of mass $M_p$
moves in a disk on a circular orbit
with radius $r_p$. The background or initial 
surface density and sound speed $\Sigma_0(r)$ and $c_0(r)$ are assumed 
to vary independently on scales $\sim r_p$.
The values of the surface density, sound speed, 
orbital frequency, and disk 
vertical scaleheight at planet's location  
are $\Sigma_p, c_p, \Omega_p$, and $h_p$ respectively.
We have summarized brief descriptions of and references to definitions 
of the most important variables we use in Table 1.

We are only interested in
the radial evolution of the surface density and neglect azimuthal variations.
Then the general time-dependent equations of disk surface density 
evolution are the continuity and angular momentum equations (Pringle 1981): 
\begin{eqnarray}
r\frac{\partial\Sigma}{\partial t}+\frac{\partial}{\partial r}(r\Sigma v_r)=0,
\label{contin}\\
r\frac{\partial}{\partial t}(\Sigma r^2 \Omega)+\frac{\partial}{\partial r}
(r\Sigma v_r \Omega r^2)=-\frac{1}{2\pi}\left(\frac{\partial F}{\partial r}+
\frac{\partial G}{\partial r}\right),~~~~~~~~~~G=-2\pi r^3 \nu \Sigma 
\frac{d\Omega}{dr},
\label{angmomcons}
\end{eqnarray}
where $\Sigma$ is the disk surface density, 
$v_r$ is a fluid
radial velocity, $F$ is the (azimuthally averaged) angular momentum flux 
transferred to the disk fluid from the planetary
tidal perturbations, and $G$ is the usual viscous angular momentum flux ($\nu$
is a kinematic viscosity). 

From this system one can easily find that
\begin{equation}
v_r=-\frac{1}{2\pi r\Sigma}\left[\frac{\partial}{\partial r}
(\Omega r^2)\right]^{-1}
\frac{\partial}{\partial r}(G+F).
\label{v_r}
\end{equation}
Substituting this result back into equation (\ref{contin}) one 
obtains a self-consistent equation  for the
surface density evolution:
\begin{equation}
\frac{\partial\Sigma}{\partial t}=\frac{1}{2\pi r}
\frac{\partial}{\partial r}\left\{
\left[\frac{\partial}{\partial r}(\Omega r^2)\right]^{-1}
\frac{\partial}{\partial r}(G+F)\right\}. \label{evol}
\end{equation}

Now we describe separately all the important ingredients entering this 
system of equations and all its necessary simplifications.

\begin{center}
\begin{deluxetable}{ l l l }
\tablecolumns{3}
\tablewidth{0pc}
\tablecaption{Key to important symbols.\tablenotemark{1}  
\label{table}}
\tablehead{
\colhead{Symbol}&
\colhead{Meaning}&
\colhead{Where defined}
	}
\startdata
$Q$  & Toomre's $Q$ & equation (\ref{mass_rat}) \\
$\alpha$  & viscous $\alpha$-parameter & \S \ref{evol_eq} \\
$h$  & disk vertical scaleheight & \S \ref{intro} \\
$\gamma$  & gas polytropic index & \S \ref{ang-mom} \\
$F_0$  & total angular momentum flux & equation (\ref{F_0}) \\
$\mu_{max}(Q)$  & function characterizing torque cutoff & equation (\ref{F_0}) \\
$F_J(r)$  & distance-dependent angular momentum flux & equation (\ref{F_J}) \\
$t$  & distance-like variable for the shock propagation & equation (\ref{t}) \\
$\varphi(t)$  & angular momentum damping function & equation (\ref{T_w}) \\
$l_p$  & Mach-$1$ length  & equation (\ref{t}) \\
$M_1$  & critical planetary mass for a strong wave nonlinearity & equation (\ref{M1}) \\
$M_2$  & critical planetary mass for a strong migration feedback & equation (\ref{MWH}) \\
$M_f$  & fiducial mass & equations (\ref{MWH}) \& (\ref{M_f}) \\
$x$  & radial distance from planet scaled by $l_p$ & equation (\ref{xdef}) \\
$x_{sh}$  & dimensionless shocking distance & equation (\ref{x_sh}) \\
$\zeta$  & proportionality constant in definition of $x_{sh}$  & equation (\ref{x_sh}) \\
$z$  & radial distance from planet scaled by the shocking distance & equation (\ref{z}) \\
$\sigma$  & dimensionless surface density & equation (\ref{sigma_def}) \\
$\rho_{\Sigma}$  & dimensionless surface density at planet's location & equation (\ref{sigma_def}) \\
$v_d$  & drift velocity & \S \ref{local} \& equation (\ref{finv}) \\
$\beta$  & relative difference in torques leading to migration & equations (\ref{finv}) \& (\ref{warddrift}) \\
$v$  & correction factor for drift velocity & equation (\ref{finv})\\
$\lambda_s$  & parameter characterizing strength of the feedback & equation (\ref{lambda_s}) \\
$z_0$  & cutoff distance & equation (\ref{lambda_s}) \\
$t_0$  & characteristic timescale  & equation (\ref{tau}) \\
$\lambda_\nu$  & viscous parameter & equation (\ref{lam_nu}) \\
$\lambda_t$  & tidal parameter & equation (\ref{lam_t}) \\
\enddata
\tablenotetext{1}{Symbols with subscript ``p'' imply the value of the corresponding quantity at the planet's location.}
\end{deluxetable}
\end{center}

\subsection{Angular momentum flux.}\label{ang-mom}

Goldreich \& Tremaine (GT80) have considered the
gravitational interaction of a 
gas disk with a planet embedded in it. They have demonstrated that
a perturber on a circular orbit 
exerts a torque on the disk only in the immediate vicinity
of the Lindblad resonances. Density waves launched at these locations
carry angular momentum away from the satellite
in the outer disk and towards it in the inner one. Lindblad resonances of 
order $m\sim r_p/h_p\gg 1$ located about one scaleheight
$\sim h_p$ from the planet are the 
strongest contributors to the angular momentum flux: lower order resonances 
are far from the planet and do not feel such strong tidal forcing while 
higher order ones are saturated because of the so-called ``torque cutoff''
(GT80). The total angular momentum flux across a cylinder of radius 
$r_p$ carried by all density wave harmonics
is shown to be given by the expression (GT80)
\begin{equation}
F_0=(G M_p)^2\frac{\Sigma_p r_p \Omega_p}{c_p^3}
\left\{\frac{4}{9}\mu^3_{max}(Q)[2K_0(2/3)+K_1(2/3)]^2\right\}\label{F_0},
\end{equation}
where $K_\nu$ denotes the modified Bessel function of order $\nu$ and
the function $\mu_{max}(Q)$ describes the strength of the torque cutoff.
It depends only on the disk stability parameter $Q$  mentioned previously
in equation (\ref{mass_rat}), and $\mu_{max}\approx 0.69$ for disks with $Q\gg 1$.

Beyond several $h_p$ from the planet, tidal perturbations behave basically
like sound waves (Lin \& Shu 1968). 
Their nonlinear evolution finally leads to shock formation 
(GR01). After
this happens the angular momentum flux of the density waves 
decreases because it gets transferred to the disk fluid.
Taking this into effect account Rafikov (2001) has demonstrated that the 
angular momentum flux carried by the tidal perturbations $F_J$ is
given by
\begin{eqnarray}
F_J(r)=\frac{2^{3/2}c_p^3 r_p\Sigma_p}{(\gamma+1)^2|2A(r_p)|}\Phi(M_p,t),
~~~~\mbox{where}\label{F_J}\\
\Phi(M_p,t)=\int\chi^2(M_p,t,\eta)d\eta=
\left(\frac{M_p}{M_1}\right)^2\Phi\left(M_1,\frac{M_p}{M_1}t\right),
\label{F_w}\\
t(r)=-\frac{r_p}{l_p}\int\limits^r_{r_p}\frac{\Omega(r^\prime)-\Omega_p}
{c(r^\prime) g(r^\prime)}dr^\prime,
~~~g(r)=\frac{2^{1/4}}{r_p c_p \Sigma_p^{1/2}}
\left(\frac{r\Sigma c^3}
{|\Omega-\Omega_p|}\right)^{1/2}. \label{t}
\end{eqnarray}
Here $A=(r/2)d\Omega/dr$ is Oort's $A$ constant, 
$l_p$ is the Mach-$1$ distance
[the distance from the planet at which shear velocity is equal to sound speed, 
$l_p=(2/3)h_p$ for Keplerian disks], $\gamma$ is a polytropic index,
and $M_1$ is given by equation (\ref{M1}).
The variable $t(r)$ plays the role of the distance travelled by the wave from 
the planet, and the dimensionless flux $\Phi(M_1,t)$ was calculated in 
GR01 for $M_p=M_1$. In equations (\ref{F_J})-(\ref{t}) the surface density 
$\Sigma$ and sound speed $c$ can vary arbitrarily with radius $r$.

It was also demonstrated in GR01 that the density wave shocks when $t$ reaches
\begin{equation}
t_{sh}=0.79\frac{M_1}{M_p}.\label{t_sh}
\end{equation}
In this expression we have neglected an
additional term equal to $1.89$, which takes into account the
 finite distance required
for the wake formation in the linear regime. 
One can do this only if $M_p\ll M_1$ (which is always true in this paper), 
so that the region of  linear
wake formation is separated from the region of its weakly nonlinear evolution.
 Otherwise these regions overlap and the
wave shocks before it is fully formed, 
significantly complicating the analysis.

For $t\le t_{sh}$, the total angular 
momentum flux of the waves $F_J$ is conserved and has to coincide with
$F_0$ given by the expression (\ref{F_0}); the value of $\Phi(t)$ for 
$t\le t_{sh}$ is such that this condition is fulfilled. Note also 
that although the polytropic index $\gamma$
enters equation 
(\ref{F_J}) the dimensionless flux $F_J$ does not actually depend on it
because $\Phi\propto (\gamma+1)^2$ (see GR01). Thus, we can rewrite 
equation (\ref{F_J}) as
\begin{eqnarray}
F_J(r)=F_0 ~\varphi\left(\frac{M_p}{M_1}t(r)\right),~~~\mbox{where}\nonumber\\
\varphi(t)=\Phi(M_p,t)/\Phi(M_p,t_{sh}),\label{T_w}
\end{eqnarray}
The function $\varphi(t)$ describes the damping after the 
shock is formed and is such
that $\varphi(t)=1$ when $t\la t_{sh}$, $d\varphi(t)/dt=0$ when $t=t_{sh}$, and
$\varphi(\infty)=0$. 
Its behavior is shown in Fig. \ref{fig:varphi}. In this paper
 instead of the true angular momentum flux 
dependence calculated by GR01  
we will be using for numerical convenience a  
simple analytical fit satisfying
all of the above mentioned  conditions and described 
in Appendix \ref{ap1}.

From the conservation of the total angular momentum 
(assuming a steady state) it follows that the corresponding 
angular momentum flux of the disk fluid is
\begin{equation}
F(r)=F_0 \left[1-\varphi\left(\frac{M_p}{M_1}t(r)\right)\right].
\label{F-fluid}
\end{equation}
This expression together with equations (\ref{F_J})-(\ref{t}) is to be substituted 
into equation (\ref{evol}) and then the 
surface density evolution could be calculated self-consistently
because the dependence of $F$ on $\Sigma$ is taken explicitly 
into consideration by equation (\ref{t}).

\subsection{Local approximation.}\label{local}

Using equations (\ref{F_J})-(\ref{t}) one can in principle study the 
temporal evolution
of $\Sigma$  taking into account the fact that $\Sigma_0$, $c_0$, $\Omega_0$,
etc. vary
on scales $\sim r_p$. We prefer to study the problem in a simplified setting
assuming that although gap formation is a nonlocal process it
still occurs sufficiently close to the planet. Later we will determine
the necessary conditions for this assumption to be true. 
   
This simplification allows us to set $c=c_p$, $r=r_p$,
and to use the shearing sheet approximation to represent 
the background fluid flow. 
The only disk quantity which is assumed to vary in our 
consideration is the surface density $\Sigma$.
By setting $c=c_p$ we imply that rapid 
variations  of $\Sigma$ do not violate the thermal balance in the 
disk that determines $c(r)$. Here the disk temperature distribution 
is supposed to be the result of external irradiation by 
the central star and our assumption 
of constant $c$ rests upon the absence of geometrical effects such as 
the shadowing by the gap edge (Lecar \& Sasselov 1999; Dullemond et al. 2001).
 
In this ``quasilocal'' approximation,  introducing a new variable
\begin{equation}
x=\frac{r-r_p}{l_p},\label{xdef}
\end{equation}
we obtain from equation (\ref{t}) that
\begin{equation}
t(x)=\mbox{sign}(x)
2^{-1/4}\int\limits_0^x \left[\frac{\Sigma_p}{\Sigma(x^\prime)}
\right]^{1/2}
|x^{\prime}|^{3/2}dx^\prime.
\end{equation}
If $\Sigma(x)=\Sigma_p$ we reproduce the expression for $t(x)$ obtained
in GR01.

The disk-planet interaction leads to the  radial migration of the planet
(GT80; Ward 1997a) with some drift velocity $v_d$. 
We  consider the disk evolution in the coordinate system
comoving with the planet. In this case
\begin{equation}
\Sigma(r,t)\to\Sigma(r-v_d t,t) ~~~\mbox{and}~~~
\frac{\partial\Sigma}{\partial t}
\to \frac{\partial\Sigma}{\partial t}-v_d \frac{\partial\Sigma}{\partial r}.
\end{equation}
Usually it is found that migration is inward, i.e. $v_d<0$ (Ward 1986). 
Using the definitions
\begin{equation}
\sigma(r,t)\equiv\frac{\Sigma(r,t)}{\Sigma(\infty)}, ~~~ \rho_\Sigma\equiv
\frac{\Sigma_p}{\Sigma(\infty)},\label{sigma_def}
\end{equation}
one can rewrite equation (\ref{evol}) in the following form:
\begin{eqnarray}
\frac{\partial \sigma}{\partial t}-\frac{v_d}{l_p}\frac{\partial \sigma}
{\partial x}=
\frac{3\nu}{l_p^2}\frac{\partial^2 \sigma}{\partial x^2}+
\frac{F_0}{\pi\Sigma(\infty)\Omega r^2 l_p^2}\frac{\partial^2}
{\partial x^2}
\left[\varphi\left(
\mbox{sign}(x)\frac{\rho_\Sigma^{1/2}}{2^{1/4}}\frac{M_p}{M_1}\int
\limits_0^x
\frac{|x^{\prime}|^{3/2}}{\sigma^{1/2}}dx^\prime\right)\right].\label{evol1}
\end{eqnarray}
Here $\Sigma(\infty)$ 
is the surface density of the disk far away
from the planet; it is different from $\Sigma_p$ 
in the case of viscous disks ($rho_\Sigma\neq 1$, see \S \ref{viscous}). 
Then the
 ratio $\rho_\Sigma$ is not known a priori and has to be found from a
self-consistent solution of the problem.
We take $\Sigma(\infty)$ to be the same both inside and outside of 
the planet which  means that
global surface density gradients 
are neglected here as we have mentioned earlier.

GR01 has demonstrated that in the constant surface density disk,
the wave shocks after propagating a distance (in $x$)
\begin{equation}
x_{sh}\approx 1.4\left(\frac{\gamma+1}{12/5}\frac{M_p}{M_1}\right)^{-2/5}=
\zeta\left(\frac{M_p}{M_1}\right)^{-2/5}, ~~~\zeta=1.4\left(\frac{\gamma+1}
{12/5}
\right)^{-2/5}.\label{x_sh}
\end{equation}
away from the planet.
In what follows we will always assume for simplicity
that $\zeta=1.4$ thus taking 
$\gamma=7/5$ [see however Fridman \& Gor'kavyi (1999)].
In a disk with varying surface density
$x_{sh}$ is no longer the distance after which the shock forms. Still, it 
is convenient to change $x$  to a new spatial coordinate $z$ given by
\begin{equation}
z\equiv
\frac{x}{x_{sh}}=\frac{x}{\zeta} \left(\frac{M_p}{M_1}\right)^{2/5}.\label{z}
\end{equation}
In terms of $z$, the surface density in our particular problem 
varies on scales $z\sim 1$. 
Thus, the condition
for applying our ``quasilocal'' approach (see \S \ref{local})
is 
\begin{equation}
1\la x_{sh}\la r_p/h_p.\label{quazi} 
\end{equation}
We check later in \S \ref{disc} 
if this is fulfilled in realistic protoplanetary systems.

With the aid of (\ref{z}) equation (\ref{evol1}) transforms to
\begin{eqnarray}
\frac{\partial \sigma}{\partial t}-\frac{v_d}{\zeta l_p}
\left(\frac{M_p}{M_1}\right)^{2/5}
\frac{\partial \sigma}{\partial z}=
\frac{3\nu}{\zeta^2 l_p^2}\left(\frac{M_p}{M_1}\right)^{4/5}
\frac{\partial^2 \sigma}{\partial z^2}\nonumber\\
+~\frac{9}{4}\frac{F_0}{\zeta^2\pi M_f\Omega r^2}
\left(\frac{M_p}{M_1}\right)^{4/5}
\frac{\partial^2}{\partial z^2}
\left[\varphi\left(I(z)\right)\right],\label{evol2}\\
I(z)=\mbox{sign}(z)\frac{\rho_\Sigma^{1/2}\zeta^{5/2}}{2^{1/4}}\int\limits_0^z
\frac{|z^{\prime}|^{3/2}}{\sigma^{1/2}}dz^\prime,~~~
M_f=\Sigma(\infty)h^2\label{M_f}.
\end{eqnarray}

Equation (\ref{evol2}) does not have $M_p$ inside the argument of $\varphi$
and this permits one to extract an explicit dependence on the planetary mass.

\subsection{Evolution equations.}\label{evol_eq}

Equation  (\ref{evol2}) can only describe the surface density evolution when
it is supplied with the information about the behavior of the 
migration velocity $v_d$ 
as a function of $\sigma$. In Appendix \ref{drift} we demonstrate that
\begin{equation}
v_d=-\frac{2\beta F_0}{M_p r_p\Omega_p}\frac{h_p}{r_p}~ v,~~~\mbox{where}~~~
v=1-\lambda_s\left(
\int\limits^{-z_0}_{-\infty}-
\int\limits_{z_0}^\infty\right)
\frac{\rho_\Sigma^{-1}\sigma-1}{z^4}dz,\label{finv}
\end{equation}
is a correction factor for the drift velocity caused by the feedback 
from the surface density variations to
the planetary migration; also  
\begin{eqnarray}
\lambda_s\equiv\frac{3}{\beta\zeta^3\mu_{max}^3}\frac{r_p}{h_p}
\left(\frac{M_p}{M_1}\right)^{6/5} ~~~\mbox{and}~~~
z_0=(\mu_{max}x_{sh})^{-1}.\label{lambda_s}
\end{eqnarray}
In the expression for $v$ the migration feedback is represented by
second term --- it takes into account modification of the contribution from 
low-order Lindblad resonances, which is important because 
the planetary torque in the outer and inner parts of the disk
 due to 
high-order Lindblad resonances almost completely cancel. 
This means that even a small contribution from low-order 
resonances could  produce a significant effect on the migration speed.
When the feedback is absent the drift speed is determined only by 
global gradients of $\Sigma_0$ and $c_0$ represented by factor $\beta$
in equation (\ref{finv}) (see Appendix \ref{drift}).

We will see later that in interesting cases the planet produces variations
of $\sigma\sim 1$ and  $\sigma>1$ in the direction of migration and $\sigma<1$ 
in the opposite direction: a surface density increase forms in the disk 
{\it in front} of the planet, and a depression appears {\it behind} it.
Thus density perturbations on both sides of the planet
 work in one direction --- to enhance the effect of the 
disk in front of the drifting planet and decrease it
behind the planet [neglecting 
for a moment the fact that 
factor $\rho_\Sigma$ in equation (\ref{finv}) could be different from unity]. 
They do not cancel each other
(like in the case of migration due to variations of $\Sigma_0$) 
and produce $O(1)$ effect. This means that planetary-driven surface density 
perturbations tend to 
{\it stall} migration. However, these variations are
strongest at distances $\sim x_{sh} h_p\gg h_p$ from the planet, where torque 
generation is weak, thus they need to be large to compete with usual
drift due to weak
global gradients (for which cancellation effects are important).

Using this and the results of previous sections we
 can rewrite equations (\ref{evol2})-(\ref{M_f})
to find the following form 
of the evolution equation :
\begin{eqnarray}
\rho_\Sigma^{-1}\frac{\partial \sigma}{\partial \tau}+v
\frac{\partial \sigma}{\partial z}=\rho^{-1}_\Sigma\lambda_\nu
\frac{\partial^2 \sigma}{\partial z^2}+\lambda_t 
\frac{\partial^2}{\partial z^2}\left[\varphi
\left(I(z)\right)\right],
\label{main}\\
\tau\equiv\frac{t}{t_0}, ~~~t_0\equiv\Omega^{-1}\times\frac{3\zeta}{4\beta} 
\left\{\frac{4}{9}\mu^3_{max}(Q)[2K_0(2/3)+K_1(2/3)]^2\right\}^{-1}
\left(\frac{M_1}{M_p}\right)^{7/5}\frac{M_1}{M_f}\frac{r_p}{h_p},\label{tau}\\
\lambda_\nu\equiv\alpha~\frac{81}{16\zeta\beta}
\left\{\frac{4}{9}\mu^3_{max}(Q)[2K_0(2/3)+K_1(2/3)]^2\right\}^{-1}
\left(\frac{M_p}{M_1}\right)^{-3/5}\frac{M_1}{M_f}\frac{r_p}{h_p}
\label{lam_nu},\\
\lambda_t\equiv\frac{3}{4\zeta\beta}
\left(\frac{M_p}{M_1}\right)^{7/5}\frac{M_1}{M_f}\label{lam_t}.
\end{eqnarray}
Here we have parameterized viscosity using the usual $\alpha$-prescription:
$\nu=\alpha h c$ and the correction factor for drift velocity $v$ is given by
equation (\ref{finv}).

We will often use the numerical form of the parameters entering this system,
namely
\begin{eqnarray}
\lambda_t\approx 0.16~Q\left(\frac{M_p}{M_1}\right)^{7/5},~~~
\lambda_s\approx 0.48~\frac{r_p}{h_p}\left(\frac{M_p}{M_1}\right)^{6/5},
\label{lambdas}\\
\lambda_\nu\approx 1.2~\alpha~ Q~
\frac{r_p}{h_p}\left(\frac{M_p}{M_1}\right)^{-3/5},~~~
t_0=\Omega^{-1}\times 0.3~Q~
\frac{r_p}{h_p}\left(\frac{M_p}{M_1}\right)^{-7/5}.\label{t_0}
\end{eqnarray}
where we have used equation (\ref{mass_rat}) and our 
calculation assumes that $Q\gg 1$, $\beta\approx 7$, and 
$\zeta\approx 1.4$.

Equation (\ref{main}) describes the evolution of the disk surface density
and its solution depends on 
 $5$ dimensionless parameters:
$\lambda_\nu,\lambda_t,\lambda_s$, $t_0$, and $z_0$ which themselves are 
combinations of $h_p/r_p$, $M_p/M_1$, $\alpha$, and $Q$.
The parameter $\lambda_t$ describes the effect of the planetary tidal torques
on the surface density distribution, while $\lambda_\nu$ tells us how 
strong the viscosity is. Obviously, for the gap to form the second term 
in the r.h.s. of (\ref{main}) must be larger than the viscous one.
The feedback which planetary migration receives from the surface density 
perturbations is described by the parameter $\lambda_s$ defined in equation 
(\ref{lambda_s}) and it is strongest for large $\lambda_s$. 
Feedback also depends on the value of $z_0$ [equation (\ref{lambda_s})] which
represents the effect of the torque cutoff.

To understand better the meaning of these parameters let us consider the 
time $\Delta t_d$ which it takes for a planet to drift across a gap with 
the width $\Delta r$. Obviously, $\Delta t_d=\Delta r/v_d$, where again 
$v_d$ is a drift velocity. Using equations (\ref{F_0}) and (\ref{finv}) one
can easily obtain that
\begin{eqnarray}
v_d\sim c~\frac{\beta}{Q}\frac{h_p}{r_p}\frac{M_p}{M_1}, ~~~~
\mbox{and}~~~~\Delta t_d\sim
\frac{\Delta r}{c}\frac{Q}{\beta}\frac{r_p}{h_p}\frac{M_1}{M_p}\label{dt_d}.
\end{eqnarray}
Time needed for viscous diffusion to fill the gap is clearly
\begin{eqnarray}
\Delta t_\nu\sim\frac{(\Delta r)^2}{\nu}\sim
\frac{1}{\Omega_p}\frac{1}{\alpha}\left(\frac{\Delta r}{h_p}\right)^2.
\label{dt_nu}
\end{eqnarray}
Finally, for the tidal torques to clear out a gap with width $\Delta r$
an angular momentum $H=\Sigma\Omega\left(r\Delta r\right)^2$ must be 
supplied to the gas (see GT80). If we assume that angular momentum flux
carried by the density waves is damped on the characteristic length 
$\sim \Delta r$, then the timescale for gap opening is
\begin{eqnarray}
\Delta t_{open}\sim\frac{H}{F_0}\sim
\frac{1}{\Omega}\frac{r_p}{h_p}\left(\frac{\Delta r}{h_p}\right)^2
\left(\frac{M_1}{M_p}\right)^2.
\label{dt_open}
\end{eqnarray}

In our particular problem the characteristic gap width and the wave 
damping distance are the same and given by  
$\Delta r\sim h_p x_{sh}=\zeta h_p\left(M_p/M_1\right)^{-2/5}$ [see equation 
(\ref{x_sh})]. Then, substituting this into equation (\ref{dt_d}) we obtain that
$t_0\sim \Delta t_d$, i.e. time $t_0$ is a typical timescale for a gas drift 
through the gap caused by the planetary migration (in planet's reference 
frame). One can see that 
$t_0\gg \Omega^{-1}$ for $M_p\ll M_1$ which justifies our use of 
azimuthally averaged quantities if $\lambda_t\sim 1$ 
[because $t_0/\lambda_t$ is
a characteristic timescale for the gap opening, see equation (\ref{main})].
Also for this choice of $\Delta r$ one can easily find
that $\lambda_t\sim \Delta t_d/\Delta t_{open}$, that is 
drift through the gap cannot replenish material repelled by planetary torques 
when $\lambda_t\ga 1$, and gap could be opened if viscosity is absent.
We will further confirm this conclusion in \S \ref{visc-less}.
Likewise, one can see that $\lambda_\nu\sim \Delta t_d/\Delta t_\nu$
which means that viscous diffusion is more important for the surface density 
evolution than effects of the migration when $\lambda_\nu\ga 1$. It is 
also clear that for tidal torques to overcome viscosity and open a gap in
a viscous disk one needs $\Delta t_{open}\la\Delta t_\nu$, or
$\lambda_t\ga\lambda_\nu$ (see \S \ref{viscous}).

\section{Kinematic wave solution}\label{kin_wave}

It is possible that 
for some values of parameters $\lambda_\nu,\lambda_t,\lambda_s$
solution for the surface density 
does not depend on time-like variable $\tau$. In this case, the
profile of $\sigma$ is stationary in the reference frame migrating with 
the planet.
Although planetary gravitational torques tend to repel the disk fluid,
planet is mobile enough to migrate through the forming gap
and stop its development. 
When such a time-independent solution is no longer possible, planet migration
stops and surface density evolves with time to form a gap in the disk. 

To study such kinematic wave solutions we take 
$\partial\sigma/\partial \tau=0$ in equation (\ref{main}). Integrating 
resulting equation  with respect to $z$ and using the fact that at infinity
$\sigma=1$, $\partial\sigma/\partial z=0$ and planetary torques are absent
we obtain that
\begin{eqnarray}
v(\sigma-1)=\rho^{-1}_\Sigma\lambda_\nu
\frac{\partial \sigma}{\partial z}+\lambda_t
\frac{\partial}{\partial z}\left[\varphi
\left(I(z)\right)\right],~~~
I(z)=\mbox{sign}(z)\frac{\rho_\Sigma^{1/2}\zeta^{5/2}}{2^{1/4}}\int\limits_0^z
\frac{|z^{\prime}|^{3/2}}{\sigma^{1/2}}dz^\prime,
\label{kin_main}
\end{eqnarray}
with $v$ given by (\ref{finv}).

WH89 used the same approach when studying gap formation with instantaneous 
density wave damping. In their case  $\varphi$
varies on scales $\sim h_p$ (wake generation requires a couple 
of disk scale lenghts to complete), i.e. $d\varphi/dz\sim x_{sh}$ 
in our notation. 
Also, to significantly affect the migration
speed one only needs surface density variations $\Delta\sigma\sim
h_p/r_p$ to be produced. Combined with the definition  
(\ref{lam_t}) this gives 
$\sim (h_p/r_p)M_f$ as a value of their critical mass, in agreement with
equation (\ref{MWH}). In our case this mass is larger because
wave damping occurs at greater distance from the planet.  

We now explore separately cases of purely 
inviscid disks and disks with nonzero 
viscosity. 

\subsection{Discs with $\nu=0$}\label{visc-less}

If the viscosity in the disk is absent equation (\ref{kin_main}) simplifies to
\begin{eqnarray}
\sigma^{1/2}(\sigma-1)=\mbox{sign}(z)
\frac{\rho_\Sigma^{1/2}\zeta^{5/2}}{2^{1/4}v}\lambda_t
~|z|^{3/2}\varphi^\prime
\left(I(z)\right),
\label{kin_less}
\end{eqnarray}
where $\varphi^\prime\left(I\right)\equiv d\varphi\left(I\right)/dI$.

\begin{figure}
\vspace{10.cm}
\includegraphics{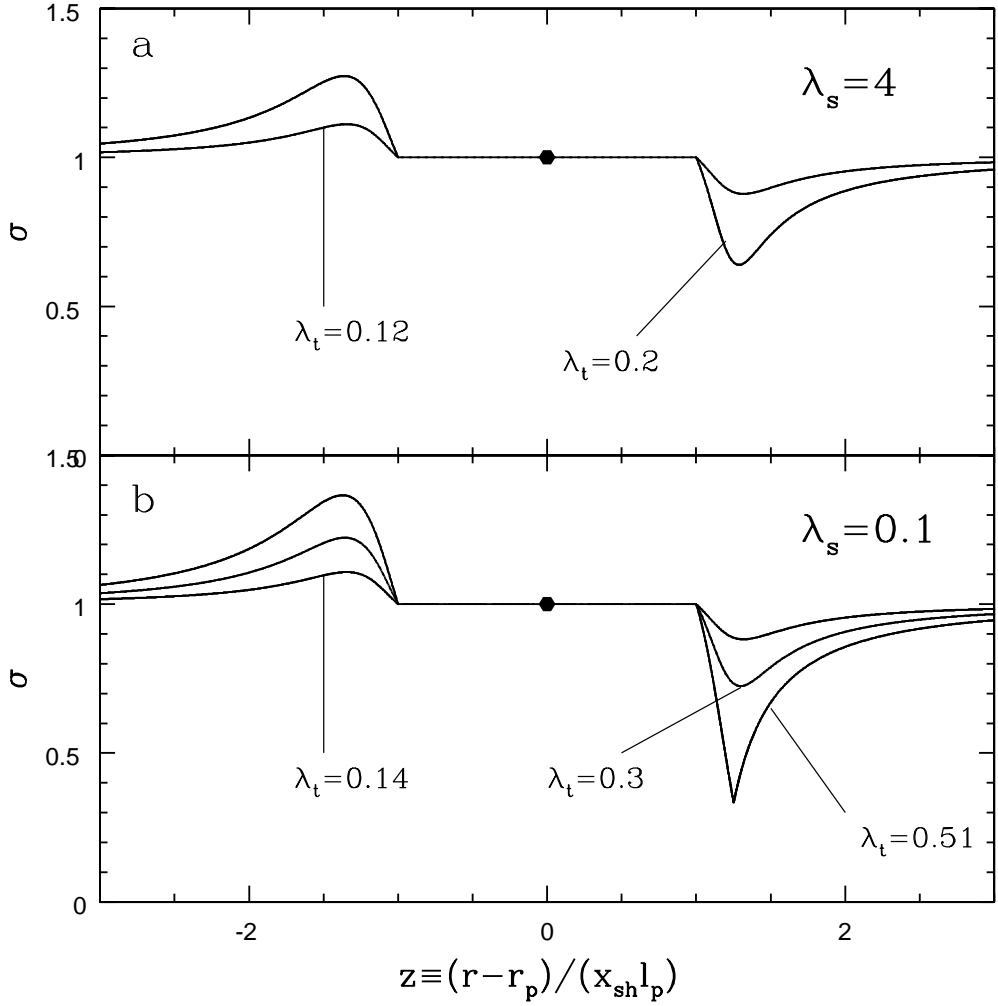}
\caption{
Surface density profiles for several values of $\lambda_t$ and $\lambda_s$
characterizing the strength of the planet's gravity and migration feedback.
Surface density normalized by its value at infinity is plotted as a function
of the distance from the perturber (normalized by the shocking distance
$x_{sh}l_p$).
Black dot denotes the position of the planet; central star 
and direction of migration are to the left.
}
\label{fig:profiles}
\end{figure}

It follows from the properties of $\varphi^\prime(t)$ that $\sigma=1$ for 
$t<t_{sh}$, i.e. for $|z|<1$, meaning that $\rho_\Sigma=1$
(see Appendix \ref{ap1}). This implies that in inviscid disk
nothing happens with the disk fluid until tidal wave shocks. 
After that, for $|z|>1$, $\varphi^\prime<0$ suggesting that
for inward migration 
 $\sigma>1$ if $z<0$ (inner disk) and $\sigma<1$ if $z>0$ 
(outer disk) just like we mentioned before. The effect of this is
to slow down the migration.
We can 
rewrite (\ref{finv}) as
\begin{equation}
v=1-\lambda_s\left(
\int\limits^{-1}_{-\infty}-
\int\limits_{1}^\infty\right)
\frac{\sigma-1}{z^4}dz.\label{fin_less}
\end{equation}
Since integrals are $O(1)$, one needs $\lambda_s\ga 1$ for the drift
velocity to
be significantly affected. The solution of the system 
(\ref{kin_less})-(\ref{fin_less}) depends on $2$ parameters only, 
$\lambda_t$ and $\lambda_s$. 

In Fig. \ref{fig:profiles} we plot the surface density profiles in the vicinity
of the planet for several values of $\lambda_t$ and $\lambda_s$. Surface 
density increases in front of the migrating planet because there are two 
fluxes of the disk fluid there --- one caused by the migration, another by 
the planetary repulsion --- and they are converging causing a
surface density pileup. Behind the moving planet these fluxes are in the same
direction --- away from the planet. For large enough planetary mass repulsion
removes matter from the the disk behind the
planet so rapidly that a corresponding surface density decrease 
cannot be replenished by the flux due to the planetary migration 
(this flux grows with $M_p$ slower than the one due to the repulsion).
As a result a gap is carved out in the disk behind the moving planet.

To describe this process quantitatively let us notice
that r.h.s. of equation (\ref{kin_less})  depends on $\sigma$ only 
through an integral over $z$. For $z>0$ r.h.s. of (\ref{kin_less}) varies from 
\begin{equation}
f_{min}=\varphi_{min}^\prime 
\frac{\zeta^{5/2}}{2^{1/4}v}\lambda_t
~z_{min}^{3/2}
\end{equation}
to $0$, where $z_{min}$ is the value of $z$ for which  $\varphi^\prime\left(
I(z)\right)$ 
reaches minimum, and $\varphi_{min}^\prime=-0.273$
as described in Appendix \ref{ap1}. At the same time 
l.h.s. of (\ref{kin_less}) varies from $-2\times 3^{-3/2}$ to $0$ (it is 
minimized at $t=1/3$). Thus, the solution of (\ref{kin_less}) does not exist
if $|f_{min}|>2\times3^{-3/2}$ or if 
\begin{equation}
\lambda_t>\frac{2}{3^{3/2}} 
\frac{2^{1/4}v}{\zeta^{5/2}|\varphi_{min}^\prime|}
~z_{min}^{-3/2}.
\end{equation}
Closer inspection of equation (\ref{kin_less}) reveals that $\sigma$ develops 
an infinite derivative 
when this happens.
  This would lead to the violation of the Rayleigh's criterion 
and instability associated with this will presumably help to clear out
a gap on the outer side of the disk. 

Once the gap is formed, density waves 
launched in the outer disk are not able to propagate to infinity and to 
transfer all their angular momentum to the disk material.
Instead they reflect from the edge of a forming gap and return to the planet.
Here they could interact gravitationally with planet thus canceling
part (or all) of the wave angular momentum  launched in the 
outer direction. It
would diminish the influence of the outer disk on the migration 
and planetary drift will stall. For this to occur planet needs to absorb
only a small fraction of one-sided angular momentum $F_0$, 
about $h_p/r_p$, which 
is obvious from the discussion in \S \ref{evol_eq} and Appendix \ref{drift}.

For $\lambda_s\la 1$ it turns out that  $z_{min}\approx 1.2$ and 
 $v\approx 1$
(planet is still rapidly migrating, feedback effects are not strong
enough to slow it down) when steady-state solution for $\sigma$
is no longer possible. This 
means that gap opens when
\begin{equation}
\lambda_t\ga 0.5 ~~~\mbox{if}~~~\lambda_s\la 1\label{lim_t}.
\end{equation} 
This corresponds to planetary mass
\begin{equation}
M_p\ga M_t\approx 2.3~ Q^{-5/7} M_1~~~\mbox{if}~~~\lambda_s\la 1.\label{m_t}
\end{equation}
For stable disks ($Q\gg 1$) this criterion
 gives quite low mass at which gap forms
although larger by $\sim (r_p/h_p)Q^{2/7}$
than the one suggested by Ward \& Hourigan (WH89).

\begin{figure}
\vspace{10.cm}
\includegraphics{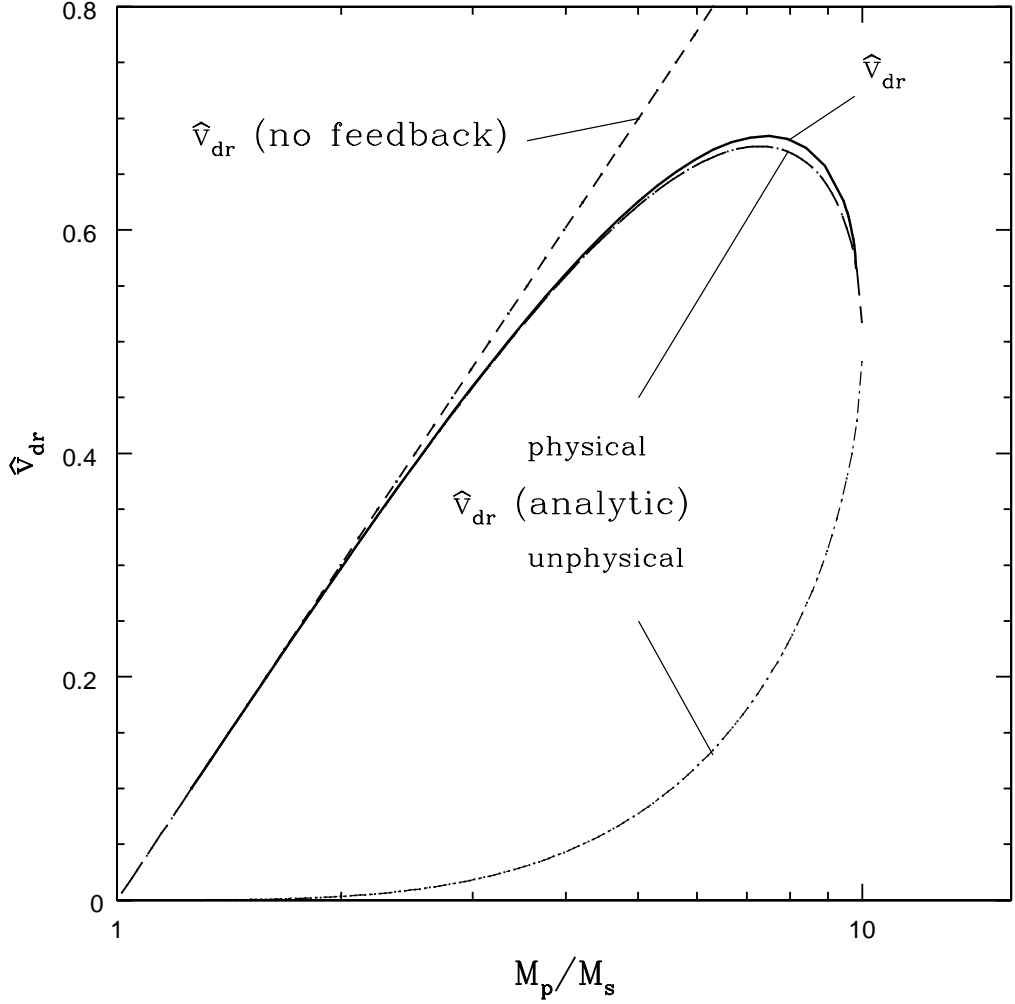}
\caption{
Drift velocity of the planet of mass $M_p$ {\it with} feedback $\hat v_{dr}$,
normalized by the drift velocity of the body 
with mass $M_s$ [see equation (\ref{mass_s})] 
{\it without} feedback, as a function of $M_p/M_s$. 
Velocity $\hat v_{dr}$ is shown by the thick solid line. Also shown 
are the 
drift velocity without feedback and analytical approximation for $\hat v_{dr}$
given by equations (\ref{v_eq}) \& (\ref{varrho}) 
(both physical and unphysical roots are displayed).}
\label{fig:drift}
\end{figure}

For $\lambda_s\ga 1$ the situation is somewhat different. 
Here the feedback from the density perturbations to the drift velocity of the
planet becomes important.
To study this region of the 
parameter space let us divide equation (\ref{kin_less}) by $z^4\sigma^{1/2}$
and integrate the result
from $-\infty$ to $-1$ and from $1$ to $\infty$, thus forming 
integrals present in r.h.s. of equation (\ref{fin_less}). Substituting
them into (\ref{fin_less}) we find that
\begin{equation}
v(1-v)=\lambda_s\lambda_t\varrho(\lambda_s,\lambda_t), ~~~\mbox{where}~~~
\varrho(\lambda_s,\lambda_t)=-\frac{\zeta^{5/2}}{2^{1/4}}\left(
\int\limits^{-1}_{-\infty}+
\int\limits_{1}^\infty\right)\frac{\varphi^\prime\left(I(z)\right)}
{\sigma^{1/2}|z|^{5/2}}dz.
\label{v_eq}
\end{equation}
It turns out that function $\varrho(\lambda_s,\lambda_t)$ 
depends on $\lambda_s,
\lambda_t$ only very weakly [which is reasonable because r.h.s.
of (\ref{v_eq}) does not depend on the rapidly varying factor $\sigma-1$]. 
We confirmed this numerically and found that
\begin{equation}
\varrho(\lambda_s,\lambda_t)\approx 0.31.\label{varrho}
\end{equation}
Since l.h.s. of equation (\ref{v_eq})
attains its maximum equal to $1/4$ at $v=1/2$, 
no solution of equation (\ref{v_eq}) for $v$ exists when 
\begin{equation}
\lambda_s\lambda_t>\frac{1}{4\varrho(\lambda_s,\lambda_t)}
~~~\mbox{if}~~~\lambda_s\ga 1.\label{lim_s}
\end{equation}

\begin{figure}
\vspace{10.cm}
\includegraphics{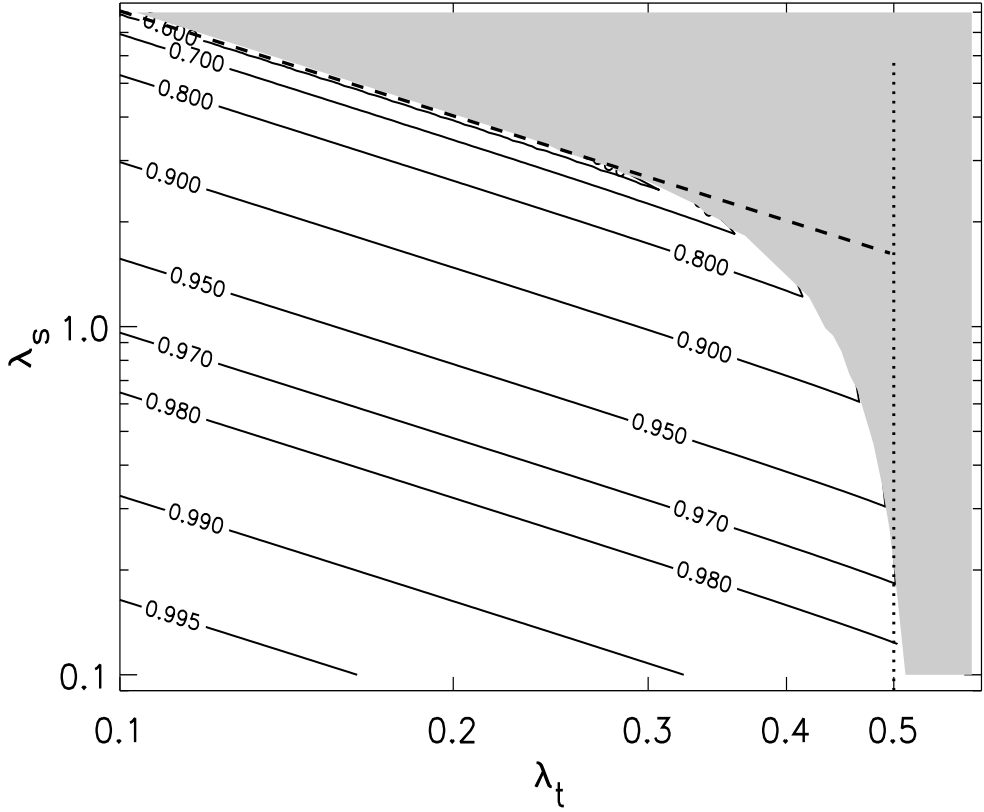}
\caption{
Contour plot of the correction factor $v$
for the drift velocity of the planet [see equation (\ref{fin_less})] ,
as a function of parameters $\lambda_t$ and $\lambda_s$. 
Shaded region corresponds to the part of the parameter space where kinematic
wave solution is not possible, migration stalls and gap opens. 
Vertical dotted line shows the 
restriction given by equation (\ref{lim_t}). 
Dashed line displays the limitation given by equation (\ref{lim_s}), 
important when the feedback is strong. 
One can see a very good agreement between the simple prescription given by 
equation (\ref{mass_lim}) and actual numerical calculation.}
\label{fig:contour}
\end{figure}

From the condition (\ref{lim_s}) and definitions (\ref{lambda_s}) \& 
(\ref{lam_t}) we can find that the feedback stops migration when 
\begin{equation}
M_p>M_s\approx5.8~\left(Q^{-1}\frac{h_p}{r_p}\right)^{5/13}M_1
~~~\mbox{if}~~~\lambda_s\ga 1.\label{mass_s}
\end{equation}

In Fig. \ref{fig:drift} we show the dependence of the drift velocity
on the mass of a perturber in the regime of strong  migration feedback. 
Drift velocity is normalized by the drift  speed which 
a body of mass $M_s$ [equation (\ref{mass_s})]
would have if the feedback were absent (straight 
dashed line on this Figure). 
Analytical solution of equation (\ref{v_eq}) is also displayed and 
is in good agreement with numerical result.  
One can easily see that $d\hat v_{dr}/dM_p=-\infty$ when $M_p=M_s$. 
This situation is analogous to what WH89 have found in their
inertial limit calculations in the case of instantaneous damping.
When a solution for the drift velocity does not exist any more planetary 
torques cannot
support a constant flux of disk material seen in planet's frame and
time dependent evolution commences leading to a gap formation and stalling the
migration. Lin \& Papaloizou (1986) and WH89 followed this process 
with time-dependent one-dimensional numerical calculations.

The behavior of the drift velocity correction factor $v$ 
is shown in Fig. \ref{fig:contour}.
One can see the cutoff of the 
possible kinematic wave solutions caused by the migration feedback
[predicted by equation (\ref{lim_s})]
on top of the plot where $v$ rapidly diminishes.
For small $\lambda_s$ the criterion given by (\ref{lim_t}) 
becomes important. The agreement 
between our simple analytical considerations and numerical results is very
good.

We can unite the gap formation conditions  
(\ref{m_t}) and (\ref{mass_s}) in a single criterion on
the limiting mass of the perturber:
\begin{equation}
\frac{M_p}{M_1}>\mbox{min}\left[2.3~ Q^{-5/7},~
5.8~\left(Q^{-1}\frac{h_p}{r_p}\right)^{5/13}\right].\label{mass_lim}
\end{equation}
Because $Q\gg 1$ we can immediately see from equation (\ref{mass_lim}) that 
gap could be formed by a planet with a mass significantly smaller than $M_1$. 

Since our analysis has only exploited the most basic features of
the damping function $\varphi(t)$ which are well reproduced by our simple 
fitting expression (\ref{match}) we conclude that the results obtained here
would be basically the same if one were to use the exact damping function
calculated in GR01.

\subsection{Viscous discs.}\label{viscous}

In the case of nonzero background viscosity present in the disk one faces a 
more complicated situation than in inviscid disk. Now the solution of 
kinematic wave equation (\ref{kin_main}) depends on $4$ parameters: $\lambda_t$,
$\lambda_s$, which we had before, $\lambda_\nu$, 
which determines the effect of the viscosity
on the solution, and $x_{sh}$, which affects drift velocity feedback. Parameter
$x_{sh}$ did not appear in the inviscid case because $\sigma$ was 
exactly $1$ near the planet in that case. Viscosity changes 
this picture because now
$\sigma$ varies near the planet, and this is important for
the feedback, because tidal forcing is very strong in the immediate vicinity 
of the planet [see equation (\ref{GT})].

In the case $\nu\neq 0$ one
can integrate equation (\ref{kin_main}) once to obtain
\begin{eqnarray}
\sigma -1 =\rho_\Sigma\frac{\lambda_t}{\lambda_\nu}\left\{
\frac{v\rho_\Sigma}{\lambda_t}\int\limits^\infty_z\exp\left[
-\frac{v\rho_\Sigma}{\lambda_\nu}(z-z^\prime)\right]\varphi
\left(I(z^\prime)\right)dz^\prime-\varphi\left(I(z)\right)\right\},
\end{eqnarray}
or
\begin{eqnarray}
\sigma -1 =\rho_\Sigma\frac{\lambda_t}{\lambda_\nu}
\int\limits^\infty_0\left[\varphi\left(I\left(z+s
\frac{\lambda_\nu}{v \rho_\Sigma}\right)\right)-\varphi\left(I(z)\right)\right]
\exp(-s)ds.\label{visceq}
\end{eqnarray}
If one integrates r.h.s. of (\ref{visceq}) by parts and takes the limit
$\lambda_\nu\to 0$, equation (\ref{kin_less}) is easily  recovered. 

As we have mentioned before, factor
$\rho_\Sigma$ in equation (\ref{visceq}) is not equal to unity any more. 
It should be determined from
the self-consistent solution of equation (\ref{visceq}) instead. 
To solve this equation 
we employ an iterative technique: at each step 
 solution obtained from the 
previous iteration is substituted into the r.h.s. of (\ref{visceq}) 
to obtain a better 
approximation until this process converges. As a first trial we use
$\sigma(z)=1$ and $\rho_\Sigma=1$.

\begin{figure}
\vspace{12.cm}
\includegraphics{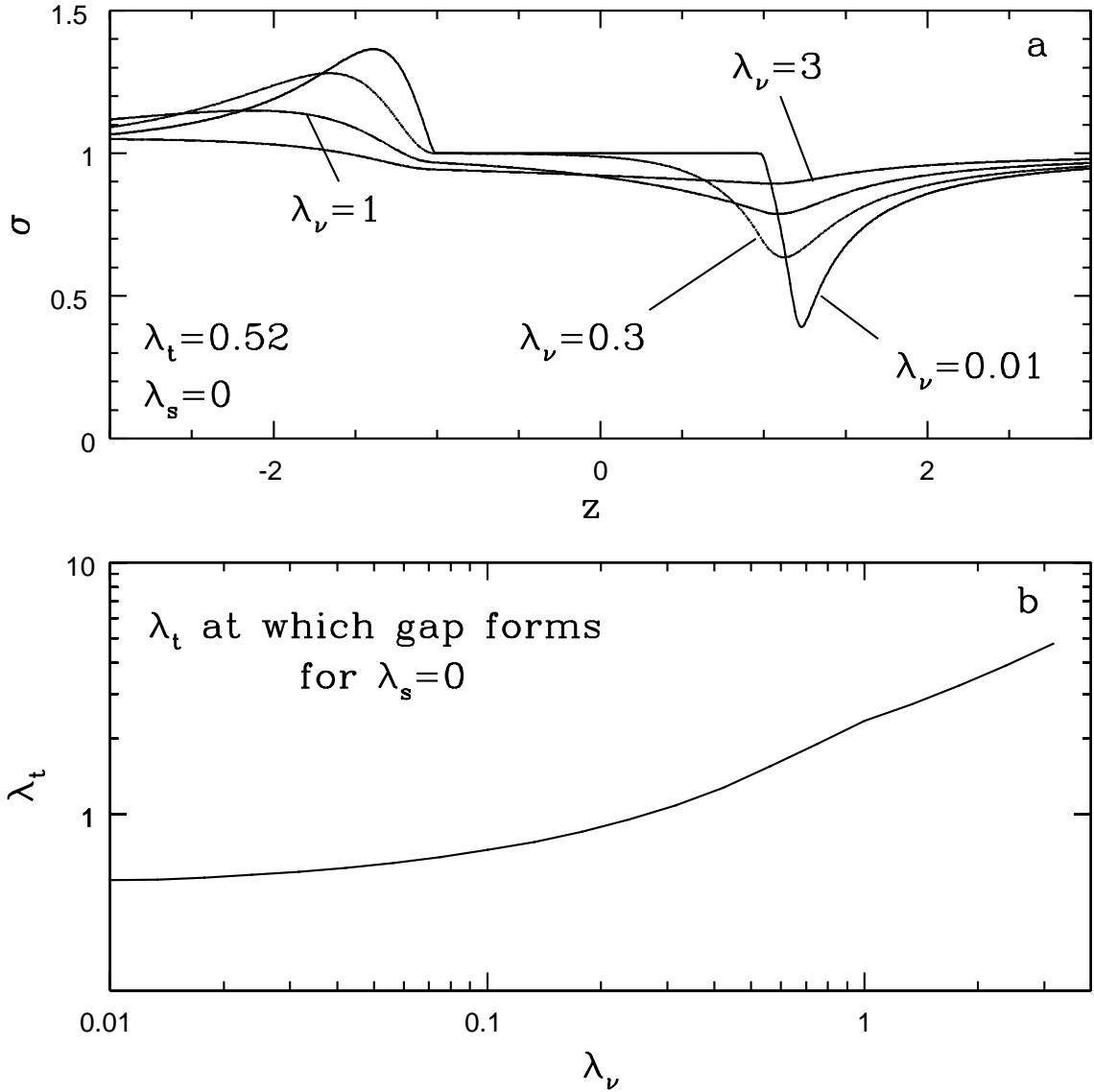}
\caption{({\it top}) Plot of the surface density as a function of distance
from the perturber for different values of viscous parameter $\lambda_\nu$:
0.01, 0.3, 1, 3. One can see how the viscosity smooths out density 
inhomogeneities in the disk and produces variations of $\sigma$
in the region $|z|\la 1$, leading to a stronger drift velocity feedback.
({\it bottom}) Maximum possible $\lambda_t$ for which planet is able to
migrate without opening a gap as a function of $\lambda_\nu$. It is assumed 
that feedback is absent: $\lambda_s=0$. One can see that viscosity 
tends to inhibit gap formation.
}
\label{fig:threshold}
\end{figure}

In Fig. \ref{fig:threshold}a we display several surface density profiles
for different $\lambda_\nu$ and fixed $\lambda_t=0.52$ 
(gap-forming value of $\lambda_t$ in the inviscid disk without the 
migration 
feedback) and $\lambda_s=0$ (thus assuming that migration feedback is absent).
For small $\lambda_\nu$ profile is almost the same as in the inviscid 
disk; one can show that $(\sigma-1)\sim \exp(-\lambda_\nu^{-1})$ 
for small $z$. As $\lambda_\nu$ grows and becomes comparable with $\lambda_t$
perturbations of $\sigma$ near $z=0$ become significant which has
profound implications for the migration feedback if $\lambda_s\neq 0$.
However, as $\lambda_\nu\ga \lambda_t$ viscous diffusion
 becomes so strong that 
surface density inhomogeneities in the vicinity of the planet become
very small (as in the case $\lambda_\nu=3$ in Fig. \ref{fig:threshold}a).

It turns out that in the viscous case (exactly like in the inviscid one) 
strong tidal perturbations can open a gap in the disk even without the   
drift velocity feedback. The presence of the viscosity only changes 
the threshold value of the parameter $\lambda_t$ so that it increases
when $\lambda_\nu$ (and, correspondingly, viscosity) grows. 
The dependence of this
limiting $\lambda_t$ on $\lambda_\nu$ for $\lambda_s=0$ 
(no feedback) is shown on Fig. 
\ref{fig:threshold}b and is easy to understand: to open a gap the planet 
has to overcome not only the drift of the fresh disk material into a 
forming gap behind the planet
but also the diffusion due to the viscous stresses which tends to 
smooth any inhomogeneities in $\sigma$ and fill the gap. Thus,  
larger $\lambda_\nu$ require larger $\lambda_t$ to open a gap.

\begin{figure}
\vspace{12.cm}
\includegraphics{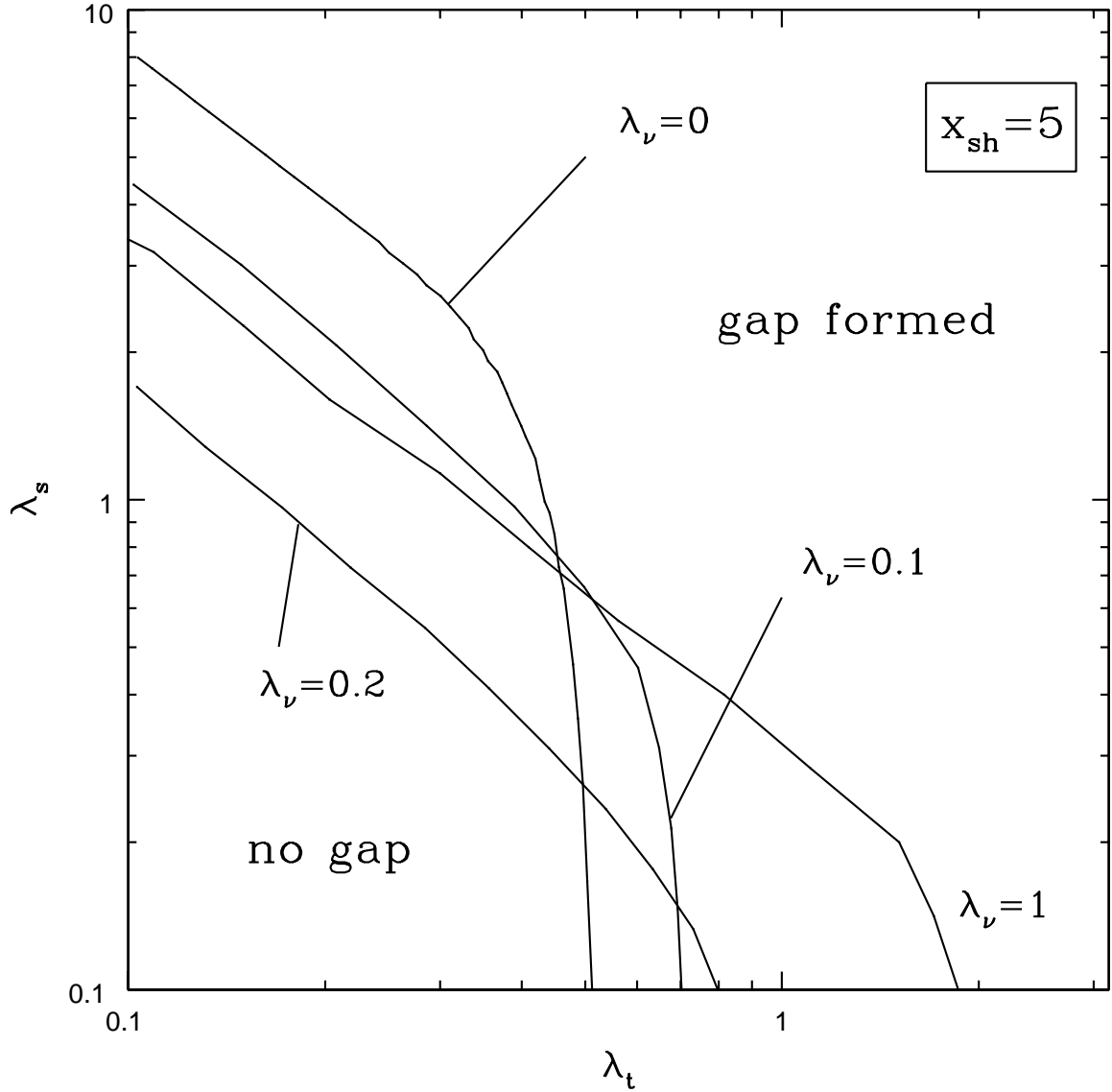}
\caption{
Boundary of the region in which planet is able to migrate without opening 
a gap for different values of the viscous parameter $\lambda_\nu$ in the
$\lambda_t-\lambda_s$ plane. Planet migrates when $\lambda_t$ and $\lambda_s$ 
are small and gap is opened for large $\lambda_t$ and $\lambda_s$. It is 
assumed that $x_{sh}=5$ is constant for all the planets (meaning that 
$M_p/M_1$ is fixed).
}
\label{fig:lam_regions}
\end{figure}

When $\lambda_s>0$ drift velocity feedback from the surface density 
perturbations could become important before $\lambda_t$ reaches the 
threshold value mentioned above. In 
Fig. \ref{fig:lam_regions} we show the boundary of the region in the 
$\lambda_t - \lambda_s$ parameter space where a planet could migrate without 
opening a gap for several values of $\lambda_\nu$. As we have said before,
for small $\lambda_s$ region where the gap cannot
be cleared expands as $\lambda_\nu$ 
increases confirming results for $\lambda_s=0$ displayed in 
Fig. \ref{fig:threshold}b. For larger $\lambda_s$, however, the situation 
is more complicated: for a fixed strength of planetary torques 
(characterized by $\lambda_t$)
this region first rapidly contracts but then expands again with
increasing $\lambda_\nu$. It could be explained as follows: 
when $\lambda_\nu\ll\lambda_t$ viscosity cannot strongly change
the surface density near the planet, $\rho_\Sigma\approx 1$ and
drift velocity feedback is only slightly stronger than in the inviscid 
case. When viscosity grows so that 
$\lambda_\nu\sim\lambda_t$, surface density perturbations are strong and
viscosity is significant to sufficiently 
modify surface density in the planetary vicinity leading to a strong 
feedback:
as it follows from equations (\ref{GT}) and (\ref{finv}) 
even small deviations of $\sigma$ from
$\sigma(0)=\rho_\Sigma$ are strongly amplified 
by a factor $\sim z^{-4}$ (and $z\sim 1/x_{sh}\ll 1$) 
in the expression for the feedback correction.
This effect  rapidly lowers the upper boundary of the region where planet
could migrate and not open a gap. However, when $\lambda_\nu\gg \lambda_t$
strong viscosity  smooths out all the inhomogeneities
in $\sigma$ which planet tends to produce and this reduces the role of the
feedback --- for a fixed $\lambda_t$ 
critical $\lambda_s$ increases.
Thus, growing viscosity tends to amplify migration feedback but 
then attenuates it as $\lambda_\nu$ exceeds $\lambda_t$.

\begin{figure}
\vspace{15.cm}
\includegraphics{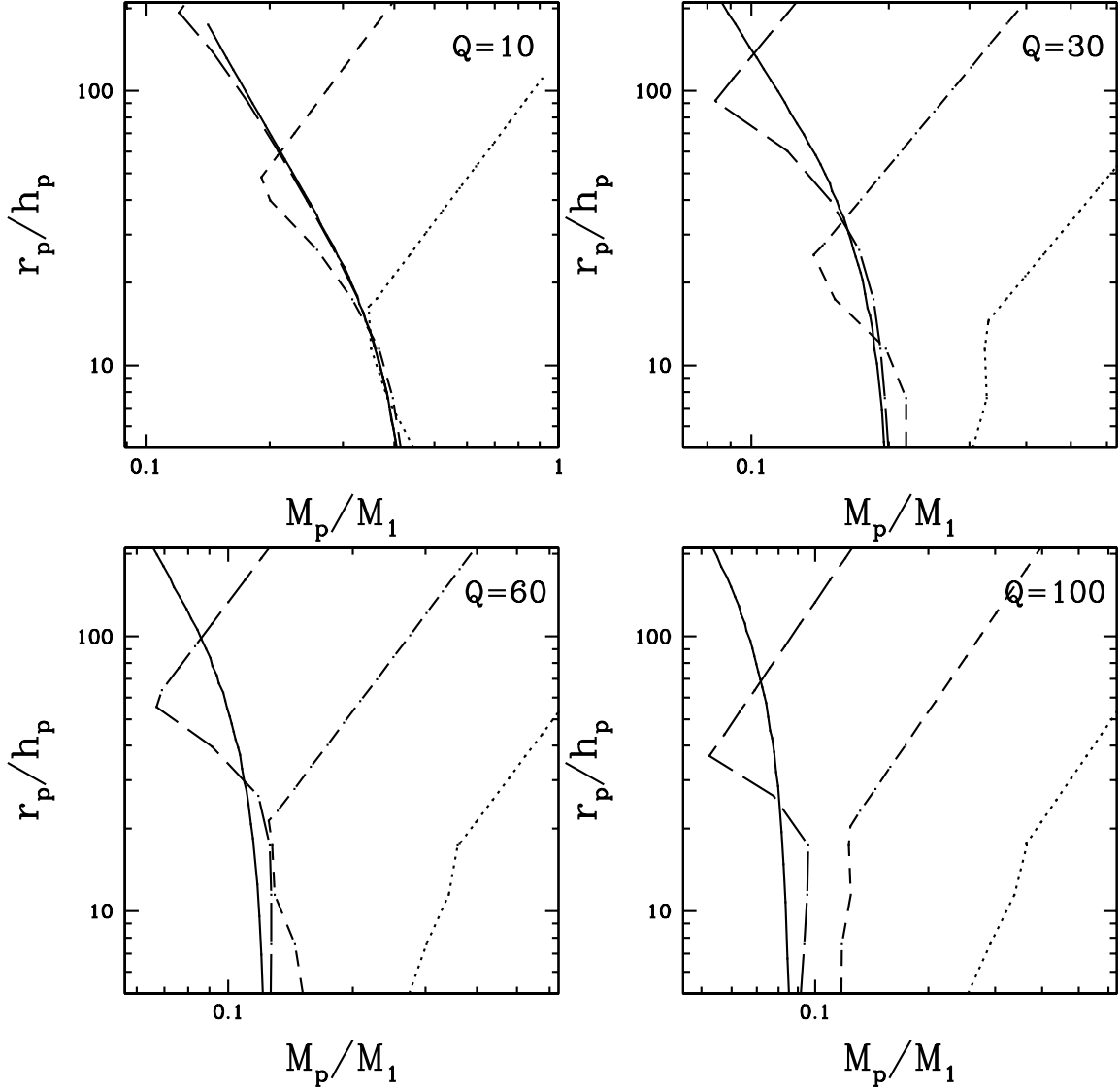}
\caption{
Boundary of the region in which planet is able to migrate without opening 
a gap for different values of dimensionless viscosity $\alpha$ 
and Toomre stability parameter $Q$ in the
$M_p/M_1 -- r_p/h_p$ plane. 
We plot boundaries for $Q=10, 30, 60, 100$ (labelled on the panels), and
$\alpha=0$ ({\it solid line}), $\alpha=10^{-5}$ ({\it long-dashed line}), 
$\alpha=10^{-4}$ ({\it short-dashed 
line}), $\alpha=10^{-3}$ ({\it dotted line}).
Gap in the disk is opened when the system is to the right of the  
corresponding boundary. 
Straight line portions of the boundaries represent restriction
$\lambda_t>\lambda_\nu$ necessary for opening a gap in the strong 
drift velocity feedback limit. 
Complicated shape of the boundary curves for $\alpha\neq 0$ is due to the 
complex interaction of the viscosity (affecting the surface density profile 
near the planet) and the drift velocity feedback.
}
\label{fig:pars_regions}
\end{figure}

In the viscous disk the absence of the steady-state solution
of equation (\ref{visceq}) caused by the feedback is not enough to
open a gap, a fact emphasized in Lin \& Papaloizou (1986) and 
WH89. Planetary torques must also
exceed viscous diffusion for this to happen. This additional 
requirement could be roughly described by a condition that
\begin{equation}
\lambda_t\ga\lambda_\nu~~~\mbox{or}~~~\frac{r_p}{h_p}\la \frac{0.133}{\alpha}
\left(\frac{M_p}{M_1}\right)^2,\label{viscond}
\end{equation}
where we have used equations (\ref{lambdas})-(\ref{t_0}). 
This is equivalent to saying that the time to open a gap by tidal torques
must be shorter than the viscous diffusion time (see \S \ref{evol_eq}). 
The condition (\ref{viscond})
coincides with the corresponding criterion used by Lin \& Papaloizou (1986) 
and WH89, but is different from the condition found in GT80 because
only gap formation by a torque exerted at a single Lindblad resonance was 
considered in GT80.
In Fig. \ref{fig:pars_regions} we plot the boundary of allowed region
in the space of physical parameters: $M_p/M_1$ and $r_p/h_p$ for 
different values of
dimensionless viscosity $\alpha$ and Toomre stability parameter $Q$. 
In depicting the boundary curves we have taken into account the 
condition (\ref{viscond}) (straight line parts of the curves are due to this
restriction).

For a fixed viscosity, the planet can open a gap only if it is 
located to the right 
of the corresponding curve in Fig. \ref{fig:pars_regions}. For small 
$r_p/h_p$ (thick disks, small migration feedback) critical $M_p$ is always 
larger than in the inviscid case and grows
with increasing $\alpha$. As $r_p/h_p$ increases both $\lambda_\nu$ and
$\lambda_\nu$ grow according to equations (\ref{lambda_s}) \& (\ref{lam_nu}).
This rapidly increases the role of the migration feedback 
so that corresponding limiting $M_p$ could become smaller than in the 
inviscid case. Complicated shape of the boundary curves for different 
values of 
$\alpha$ is due to the before mentioned
complexity of the influence of the disk viscosity
on the migration feedback.

One can see from the Fig. \ref{fig:pars_regions} that for values of $r_p/h_p$
typical in protoplanetary disks ($r_p/h_p\sim 20-30$) $M_p\ll M_1$ if $\alpha$ is small 
(typically when $\alpha<10^{-4}$). One can also notice that 
it is a migration feedback
which initiates the evolution of the surface density leading to 
the gap formation once the condition (\ref{viscond}) is fulfilled.

Our consideration of the viscous case assumes that the only surface density
gradients producing viscous flux are due to the planet-driven perturbations.
However, background surface density varying on scales $\sim r_p$ also 
contributes to the viscous fluid flux. We neglected this effect completely 
when we employed our ``quasilocal approximation'' and omitted factors 
$\Sigma/r$ leaving only $d\Sigma/dr$ in our equations
(see \S \ref{local}). If we were to 
include these factors in our analysis
 a picture similar to that described in Ward (1997a)
would be found: for large viscosity there would not be an 
infinite derivative of the drift velocity with respect to $M_p$ at some point
as in the case shown in Fig. \ref{fig:drift}.
Instead $v_{dr}$ would smoothly (but rapidly)
decrease at some point as $M_p$ grows and the migration 
would change gradually from type I to type II (Ward 1997a). 
We are unable to capture this effect working in the framework
of our ``quasilocal approximation'' but this limitation cannot affect our
major conclusions concerning the gap forming planetary mass. 
Also, when the viscosity is small or when the condition (\ref{quazi})
is satisfied we should not worry about this effect at all. Using 
equations (\ref{lambdas})-(\ref{t_0}) we find that these additional viscous fluxes 
could be neglected if $r_p/h_p \gg (M_p/M_1)^{-2/5}$ which is usually
the case in realistic situations.

\section{Discussion}\label{disc}

\subsection{Limitations of the analysis.}\label{limits}

The validity of our consideration is restricted not only by the requirements
(\ref{quazi}) and $M_p\ll M_1$ but also by the applicability of our
nonlinear wave damping prescription. Its validity was extensively
discussed in GR01 and Rafikov (2001) and we will not repeat this 
discussion here. 
For small enough viscosity ($\alpha<10^{-4}$) we found (see Fig. 
\ref{fig:pars_regions}) that
$M_p/M_1\sim 0.1$.
This means that typically $t_{sh}\sim 10$ and
$r_{sh}\approx (2-3)\times h_p$ so that the condition $x_{sh}\ga 1$ is
marginally fulfilled (note also that $t_{sh}\gg t_0=1.89$ which is the 
offset value of $t$ characterizing the wake generation region, see GR01). 
Thus, a wave typically shocks after travelling only several $h_p$ from the 
planet; assumption $x_{sh}\ll r_p/h_p$ is reasonable if $r_p/h_p\ga (10-20)$ 
which is the case in MMSN model described by
equation (\ref{hayashi}) ($r_p/h_p\approx 25$ at $1$ AU).
In disks with higher viscosity, the wave will shock even closer to the planet
(see \S \ref{viscous}) meaning that initially wake is not tightly wound, 
but this can hardly change our gap formation criterion.

\subsection{Time-dependent evolution}\label{time-dep}

After the planet grows so massive  that a steady-state surface density 
distribution is no longer possible,
time-dependent evolution commences. We do not study this process here.
It is clear however that as soon as the gap opening criterion is fulfilled and 
disk viscosity is overcome by planetary torques, gap clearing starts 
behind the planet (in the part of the disk 
 opposite to the direction of migration).
As we have mentioned before this makes propagation of the density waves in
the outer disk impossible, and planet migration which is the result of the 
small difference of the torques on both sides of the disk, has to stall. 
When this happens the planet gradually repels the surrounding material, 
carving out 
a gap in both parts of the disk. Far from the planet 
gap expansion can be stopped by the disk viscosity or tidal torques 
from other planets in the system. 

This general picture of gap formation (after the corresponding gap-opening 
criterion is fulfilled) was  confirmed by the 
one-dimensional time-dependent numerical 
calculations in Lin \& Papaloizou (1986) and WH89. Of course they 
used a different gap-forming criterion because of their assumed 
instantaneous damping of the planet-generated density waves. But because
the reason for the gap formation in our
case is basically the same as in theirs --- migration 
feedback and strong tidal torques exceed the viscous spreading ---
we expect the time-dependent transition
from type I to type II migration in our setting to be similar to what these
authors have found in the local damping approximation. The only differences 
would be the critical mass 
at which this happens and the transition timescale.
Our results for the gap-opening criterion also seem to be in general agreement
 with the time-dependent two-dimensional
simulations carried out by LP93, Bryden et al. (1999),
 and Nelson et al. (2000), although it is hard to
 make a direct quantitative comparison because our 
criterion requires knowledge of a larger number of  
parameters than is quoted in these studies. 

Typical timescale for a gap to form is $t_0/\lambda_t$.
Using equation (\ref{tau}) or (\ref{t_0}), assuming $M_p=0.1 M_1$ and 
$\lambda_t\sim 1$, one can find that for MMSN parameters this
time is about $5\times 10^4~\Omega^{-1}\approx 7\times 10^3$ yr at $1$ AU
and $\sim 2\times 10^4~\Omega^{-1}\approx 4\times 10^4$ yr at $5$ AU (semimajor
axis of Jupiter's orbit). Such long timescales can put some restrictions
on the numerical schemes which would be able to check 
the validity of
our gap-opening criterion (they should also have high spatial resolution,
typically a fraction of $h_p$, and be able to follow the formation and
evolution of weak shocks in a disk, although numerical viscosity would be a 
primary concern).

\subsection{Applications}\label{apps}

Using the theory developed in \S \ref{kin_wave} we 
can determine the planetary mass
 at a particular location in a disk for which 
gap formation is expected and migration 
switches to a type II mode.
For the MMSN parameters
represented by equation (\ref{hayashi}) one obtains 
$r_p/h_p\approx 25$, $Q\approx 70$, and $M_1\approx 14~M_\oplus$ at $1$ AU. 
Using Fig. \ref{fig:pars_regions} we find that for $\alpha\la 10^{-4}$ a
gap is opened by a planet with $M_p\approx (0.12-0.15)M_1\approx
(1.5-2)~M_\oplus$. Thus, if the Earth was immersed in a gaseous disk at the
end of its formation, it probably could not open a gap  
(nonzero disk viscosity only strengthens this conclusion). 
Then type I migration would have caused it to drift
to the Sun on a timescale $\la 10^5$ yr, but this apparently did not happen. 
This result supports the usual view according to which the final 
accumulation of terrestrial planets
via the runaway coagulation of planetesimals 
 or giant impacts of Moon-sized planetary embryos
 occurred {\it after} the nebula was dispersed in the
inner Solar System. Then Earth would not have migrated at all and the 
question of its survival would not arise. 

At $5$ AU one finds that $r_p/h_p\approx 16$, $Q\approx 45$, 
and $M_1\approx 50~M_\oplus$; then critical $M_p/M_1\approx (0.15-0.2)
M_1\approx (7-9) M_\oplus$ for $\alpha\la 10^{-4}$. Thus, even 
the usual rocky core 
of Jupiter alone (without gaseous envelope) 
would likely be able to open a gap; 
this would significantly slow down its 
migration towards the Sun and leave Jupiter on its orbit far from the central 
star. For some reason this clearly did not happen 
in systems harboring extrasolar giant planets close to their parent stars,
and this issue apparently deserves further study. 
The value of the critical mass also raises another
question: how has Jupiter managed to acquire its huge gaseous mass?
It might be that the timescale for the gap formation at the Jupiter's location
is long enough (see \S \ref{time-dep}) for the planet to accrete all its 
mass during the gap-opening stage if the core instability 
(Mizuno 1980) was operating since the very beginning (core instability
and associated planetary mass growth 
 might have actually triggered the gap formation).
But obviously more detailed consideration of the gas accretion process 
is needed to definitely answer this question.

An interesting result following from this analysis is that when a 
gap is opened
in an inviscid disk, there is still some material remaining at the 
planet's orbit. This material has the form of a ribbon with radial width 
equal to 
$2h_px_{sh}$ because waves launched by the planet cannot shock prior to 
travelling a minimum necessary distance, and, thus, cannot transfer 
their angular momentum to the disk and cause its evolution. The effects of
wave action reflection from the edge of the ribbon or the presence of the 
moderate viscosity in the disk might lead to the dissipation of 
such a gas torus. But if these effects are not very strong then this ribbon 
phenomenon might have observable manifestations.

\section{Conclusions}

Using a realistic damping prescription for tidally-induced density
waves we have studied the conditions necessary for a gap to be formed in
a gas disk in the vicinity of a planet. It was shown that for
small enough planetary mass a steady-state solution for the surface 
density perturbations exists in the
reference frame migrating with the planet. Then the 
details of the surface density
distribution in the disk depend on only $4$ parameters: the aspect ratio
in the disk $h_p/r_p$, the viscosity (represented here by its 
dimensionless analog 
$\alpha$), the Toomre stability parameter of the disk $Q$, and the ratio
of the planetary mass to a fiducial mass $M_1$ defined in equation (\ref{M1}). 
Only systems with $\alpha<10^{-3}$ were considered because larger viscosity
could violate our damping prescription based on the nonlinear wave 
dissipation. We have demonstrated 
that in disks with $Q$ between $\sim 10$ 
and $\sim 100$ (which should be typical
in passive protoplanetary disks irradiated by their central stars) a
gap is opened when the planetary mass reaches $(1-10)M_\oplus$,
 depending on the disk viscosity and the planet's location in the nebula. 
Planets further away from the
central star must be more massive to repel the gas in their vicinity.
We obtained an analytical criterion  for a gap-forming 
planetary mass
in inviscid disks [see equation (\ref{mass_lim})].
Our result for this critical mass is in between the previous estimates
of this quantity 
obtained by Ward \& Hourigan (1989) and Lin \& Papaloizou (1993)
because of the different density wave 
damping function (WH89) and moderate requirements
for the nonlinearity of the wave (LP93).

The apparatus developed here for following the disk surface 
density evolution and 
studying stationary structures in the disk such as the 
kinematic wave solutions 
(see \S \ref{kin_wave}) could be generalized to other 
astrophysical problems; for example 
it could be used to study global disk evolution or to 
extend our gap formation analysis to a system 
containing several planets.

\section{Acknowledgements}

I want to express my gratitude
to Scott Tremaine and Jeremy Goodman for useful
discussions and a lot of thoughtful comments on the manuscript. 
Financial support 
provided by the  NASA grant 
NAG 5-10456 is thankfully acknowledged.

\appendix

\section{Angular momentum flux $\varphi(t)$.}\label{ap1}

GR01 have calculated the dependence of the angular momentum flux 
of shock damped density waves using the full numerical solution
for the propagation and damping of the wake. Here for simplicity 
we use a simple matching function which reproduces the main features 
of the numerical solution:
\begin{equation}
\varphi(t)=\left\{
\begin{array}{rcl}
0, ~~~t<t_{sh},\\
\left[1+\left(t/t_{sh}-1\right)^2\right]^{-1/4}, ~~~
t>t_{sh}.\\
\end{array}
\right.
\label{match}
\end{equation} 

\begin{figure}
\vspace{10.cm}
\includegraphics{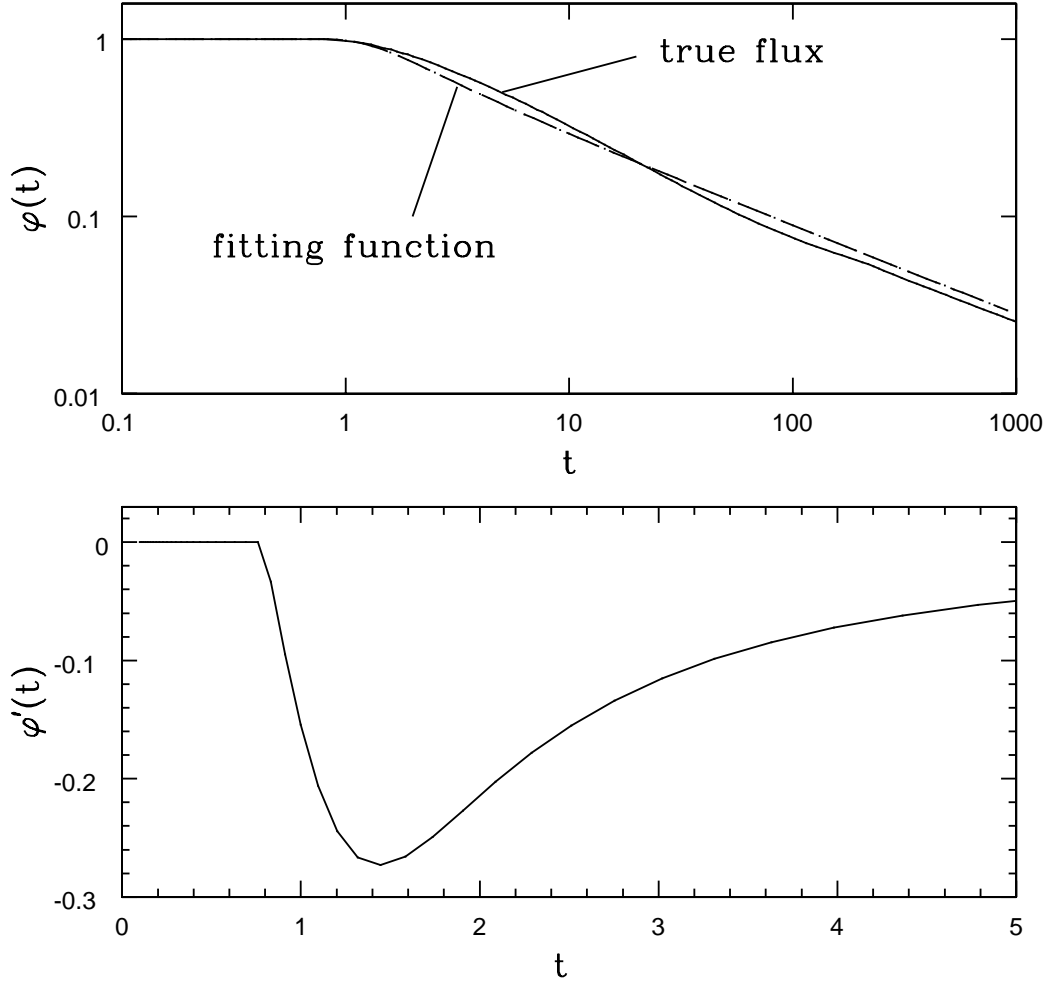}
\caption{
({\it top})
Behavior of the dimensionless angular momentum flux $\varphi$ (solid line) 
and a simple fit (dashed line) given by equation (\ref{match}). 
The flux diminishes as $t$ increases 
due to shock damping.
({\it bottom}) Plot of $\varphi^\prime(t)$ calculated using the 
fitting function
(\ref{match}).}
\label{fig:varphi}
\end{figure}

The behavior of the true solution and the matching function
is shown in Fig. \ref{fig:varphi}. For a 
planet-generated wake profile  $t_{sh}=0.79$ (neglecting the fact that
there is some distance $t_0$ over which the wake profile forms).
Our simple fit never deviates from the true $\varphi(t)$ by more
than $17\%$. Asymptotically $\varphi(t)\propto t^{-1/2}$ for $t\to \infty$,
as in the numerical solution. Also $d\varphi(t)/dt=0$ when $t=t_{sh}$.

To study the surface density profile near the planet in inviscid disks
it is important to know the behavior of $\varphi^\prime(t)=d\varphi(t)/dt$.
For our assumed matching function (\ref{match}) one can find
that 
\begin{eqnarray}
\varphi^\prime(t)\propto (t_{sh}-t), ~t-t_{sh}\ll 1,
~~~~\varphi^\prime(t)\propto -t^{-3/2}, 
~t\gg 1,~~~ \mbox{and} ~~~ \varphi^\prime(t)=0~\mbox{for}~t<t_{sh}.
\end{eqnarray}

The function $\varphi^\prime(t)$ reaches its minimum value equal to 
$\varphi^\prime_{min}=-0.273$ at $t_{min}=1.43$.

\section{Drift velocity}\label{drift}

From the global conservation of the angular momentum we find 
the velocity of  migration driven by the planetary gravitational torques: 
\begin{equation}
v_d=-\frac{2}{M_p r_p\Omega_p}\int\frac{\partial F}{\partial r}dr,\label{drvel}
\end{equation}
where the integral is taken over the whole disk. One can write 
$dF/dr=\Sigma(r) df/dr$, where $F$ is the angular momentum flux produced
by the planet per unit surface density at distance $r$. GT80 have
demonstrated that
function $df/dr$ rapidly
falls off like $|r-r_p|^{-4}$ for $|r-r_p|\ga h_p$.

Ward (1986) has shown that
$df/dr=\mbox{sign}(r-r_p)df/dr|_0[1+O(h_p/r_p)(r-r_p)/h_p]$,
where $df/dr|_0$ is an {\it even} function of $r-r_p$ and
terms of order $h_p/r_p$ arise because
of the 
asymmetries in the torque generation pattern intrinsic for Keplerian disks.
The surface density itself may vary on different scales. In our particular 
case we are interested in the variations on scales of order $r_p$ 
[represented by $\Sigma_0(r)$]
which lead
to migration in the first place, and variations on scales 
$\sim h_p x_{sh}\ll r_p$, which result from the planet-driven surface 
density evolution. 
In the vicinity of the planet
$\Sigma$ could be written in the following form:
\begin{equation}
\Sigma(r)=\Sigma_0(r)\left[1+\frac{\Sigma -\Sigma_0(r)}{\Sigma_0(r)}\right]
\approx
\Sigma_p\left[1+\frac{2 h_p}{3 r_p}\frac{d\ln\Sigma_0}{d\ln r}
x\right]
\times\left[1+\frac{\Sigma -\Sigma_p}{\Sigma_p}\right].
\end{equation} 
The replacement of $\Sigma_0(r)$ with $\Sigma_p$ in the last 
bracket assumes that variations due to planetary torques are more rapid
than background ones (because damping length is supposed to be much 
shorter than $r_p$).

Substituting this in (\ref{drvel}) and recalling that $df/dr|_0$ falls
off beyond several $h_p$ from the planet one finds that 
\begin{equation}
\int\frac{\partial F}{\partial r}dr=F_0\left[\beta\frac{h_p}{r_p}
+\frac{\Sigma_p}{F_0}\int \mbox{sign}(x)\frac{df}{dx}
\bigg|_0(\rho_\Sigma^{-1}\sigma-1)
dx\right].
\label{inttorque}
\end{equation}
Here $\beta$ is a constant of the order of unity which depends on the 
global gradients of the surface density $k=-d\ln\Sigma/d\ln r$
and temperature $l=-d\ln T/d\ln r$ in the disk. 
The first term in brackets arises as a result of the asymmetry 
in torques produced in the inner and outer parts of the disk.
Ward (1986) advocates that
for $Q=\infty$ ($Q$ is a Toomre stability parameter) 
\begin{equation} 
\beta\times 2\left\{\frac{4}{9}\mu^3_{max}(Q)[2K_0(2/3)+K_1(2/3)]^2\right\}
=6.5(1+0.06 k+1.2 l),\label{warddrift}
\end{equation}
and for $Q=2$ r.h.s. of (\ref{warddrift}) changes to
$72(1-0.19 k+0.95 l)$. It means that the surface density decreasing from
the center of the disk leads to a very weak acceleration of drift in 
$Q=\infty$ disk and slows it down for moderate values of $Q$. Further we will
assume for simplicity
that the expression in parentheses in r.h.s. of 
(\ref{warddrift}) is equal to $2$ (which corresponds to $\beta\approx 7$).

Using (\ref{inttorque}) we obtain that
\begin{equation}
v_d=-\frac{2\beta F_0}{M_p r_p\Omega_p}\frac{h_p}{r_p}~ v,~~~\mbox{where}~~~
v=1+\frac{1}{\beta}\frac{r_p}{h_p}\frac{\Sigma_p}{F_0}\int \mbox{sign}(x)
\frac{d f}{dx}\bigg|_0(\rho_\Sigma^{-1}\sigma-1)dx\label{v}\\
\end{equation}
is a correction factor for the drift velocity caused by the feedback 
from the surface density variations to
the planetary migration.

For large $x=(r-r_p)/l_p$ it could be demonstrated (GT80)
that 
\begin{equation}
\frac{df}{dx}\bigg|_0=\frac{F_0}{\Sigma_p}
\frac{3}{\mu_{max}^{3}x^4}.\label{GT}
\end{equation}
For $x\la 1$ torque rapidly decreases (``torque cutoff''). 
For simplicity we will assume that the dependence given by equation (\ref{GT})
holds true even for $x\sim 1$, then we only need to correct the cutoff value
of $x$ so that the integral of (\ref{GT}) gives us the 
right amount of the angular momentum
flux $F_0$ at infinity. 
It is done by setting $df/dx|_0=0$ for $|x|<\mu_{max}^{-1}$; 
for $|x|>\mu_{max}^{-1}$ we assume $df/dx|_0$ to be given by equation (\ref{GT}).
Using  equation (\ref{v}) we finally obtain equation  (\ref{finv}) and (\ref{lambda_s}).

Ideally, 
one should also take into account gradient of $\Sigma$ on short
damping scale in the expression for $df/dr$ as it was done by Ward (1986).
This could  produce some modification because 
surface density gradients displace the positions of the Lindblad resonances
and modify the torque cutoff. 
However, while variations of $\Sigma$ produce contribution to the drift 
$\propto\Delta \sigma$, presence of the gradients of $\Sigma$ gives rise
to the effects of order $d\sigma/dx\sim\Delta\sigma/x_{sh}$.
This implies that contribution due to the gradient of the surface density
is
$\sim 1/x_{sh}$ compared with the last term in (\ref{inttorque}) and is 
unimportant for $M_p\ll M_1$.

\end{document}